%% file: main.tex
\documentclass[journal,compsoc]{IEEEtran}
\RequirePackage{fix-cm}
\usepackage[english]{babel}
\usepackage[utf8x]{inputenc}
\usepackage[T1]{fontenc}
\usepackage{cite}

\usepackage{amsmath}
\usepackage{graphicx}
\usepackage[colorinlistoftodos]{todonotes}
\usepackage[colorlinks=true, allcolors=blue]{hyperref}
\usepackage{fancyvrb}
\usepackage[normalem]{ulem}

\input{commands.tex}

\title{The Corrective Commit Probability \\ Code Quality Metric}
\hide{
\author{Idan Amit \and Dror G. Feitelson}
\institute{Department of Computer Science\\
The Hebrew University of Jerusalem, 91904 Jerusalem, Israel\\
\email{idan.amit@mail.huji.ac.il}\\
\email{feit@cs.huji.ac.il}
}
}
\author{\anon{
\begin{tabular}{c}
Idan Amit\\
idan.amit@mail.huji.ac.il
\end{tabular} ~~~
\begin{tabular}{c}
Dror G. Feitelson \\
feit@cs.huji.ac.il
\end{tabular} \\
Department of Computer Science\\
The Hebrew University of Jerusalem, 91904 Jerusalem, Israel\\
}{[Authors anonymized]}}

\begin{document}
\maketitle

\begin{abstract}
We present a code quality metric, Corrective Commit Probability (CCP), measuring the probability that a commit reflects corrective maintenance.
We show that this metric agrees with developers’ concept of quality, informative, and stable. 
Corrective commits are identified by applying a linguistic model to the commit messages.
We compute the CCP of all large active GitHub projects (7,557 projects with 200+ commits in 2019). 
This leads to the creation of a quality scale, suggesting that the bottom 10\% of quality projects spend at least 6 times more effort on fixing bugs than the top 10\%.
Analysis of project attributes shows that lower CCP (higher quality) is associated with
smaller files,
lower coupling,
use of languages like JavaScript and C\# as opposed to PHP and C++,
fewer developers,
lower developer churn, better onboarding,
and better productivity.
Among other things these results support the ``Quality is Free'' claim, and suggest that achieving higher quality need not require higher expenses.
\hide{
Analysis based on CCP supports the negative influence of high file length and coupling on quality. 
Our results support the ``Quality is Free'' claim and suggest aiming for quality will lead to benefits in developers' churn, and productivity.
}
%\keywords{Code quality metric \and Process metric \and Corrective commits}
\end{abstract}

\section{Introduction}

It is widely agreed in the software engineering community that code quality is of the utmost importance.
This has motivated the development of several software quality models (e.g.\ \cite{boehm76,dromey95}) and standards.
Myriad tools promise developers that using them will improve the quality of their code.

But what exactly is high-quality code?
There have been many attempts to identify issues that reflect upon code quality.
These generally go under the name of ``code smells'' \cite{1999:RID:311424, 1173068}.
Yet, it is debatable whether commonly used metrics indeed reflect real quality problems \cite{alkilidar05,bavota15}.
Each smell is limited to identifying a restricted shallow type of problems.
Besides, sometime a detected smell (e.g., a long method) is actually the correct design due to other considerations.

A more general approach is to focus on an indirect assessment of the software based on the ill-effects of low quality--- and in particular, on the presence of bugs.
There is no debate that bugs are bad, especially bugs reported by customers \cite{hackbarth16}.
By focusing on actual bugs, one is relieved of the need to consider all possible sources of quality problems, also removing any dependence on the implementation of the procedure that finds them.
Moreover, approaches based on bugs can apply equally well to different programming languages and projects.

Based on these considerations, we suggest the Corrective Commit Probability (CCP, the probability that a given commit is a bug fix) as a metric of quality.

Corrective maintenance (aka fixing bugs) represents a large fraction of software development, and contributes significantly to software costs \cite{lientz78, schach03b,6191}.
But not all projects are the same: some have many more bugs than others.
The propensity for bugs, as reflected by their fixing activity, can therefore be used to represent quality.
This can be applied at various resolutions, e.g., a project, a file, or a method.
Such application can help spot entities that are bug prone, improving future bug identification \cite{Walkinshaw:2018:FRD:3239235.3239244,Kim:2007:PFC:1248820.1248881, Rahman2011BugCacheFI}.

While counting the number of bugs in code is common, disregarding a project's history can be misleading.
In the CCP metric we normalize the number of bug fixes by the total number of commits, thereby deriving the probability that a commit is a bug fix.
We focus on commits because in contemporary software development a commit is the atomic unit of work.

We identify corrective commits using a linguistic model applied to commit messages, an idea that is commonly used for defect prediction \cite{Ray:2014:LSS:2635868.2635922, 5463279,  Hall:2012:SLR:2420627.2420790}.
The linguistic-model prediction marks commit messages as corrective or not, in the spirit of Ratner et al.\ labelling functions \cite{NIPS2016_6523}.
Though our accuracy is significantly higher than previous work, such predictions are not completely accurate and therefore the model hits do not always coincide with the true corrective commits.

Given an implementation of the CCP metric, we perform a large-scale assessment of GitHub projects.
We analyze all 7,557 large active projects (defined to be those with 200+ commits in 2019, excluding redundant projects which might bias our results \cite{5069475}).
We use this, inter alia, to build the distribution of CCP, and find the quality ranking of each project relative to all others.
The significant difference in the CCP among projects is informative.
Software developers can easily know their own project's CCP.
They can thus find where their project is ranked with respect to the community.

Note that CCP provides a retrospective assessment of quality.
Being a process metric, it only applies \emph{after bugs are found}, unlike code metrics which can be applied as the code is written.
The CCP metric can be used as a research tool for the study of different software engineering issues.
A simple approach is to observe the CCP given a certain phenomenon (e.g., programming language, coupling).
For example, we show below that while investment in quality is often considered to reduce development speed, in practice the development speed is actually higher in high quality projects.

Our \textbf{main contributions} in this research are as follows:
\begin{itemize}
\item We define the Corrective Commit Probability (CCP) metric to assess the quality of code.
The metric is shown to be correlated with developers' perceptions of quality, is easy to compute, and is applicable at all granularities and regardless of programming language.
\item We develop a linguistic model to identify corrective commits, that performs significantly better than prior work and close to human level.
\item We show how to perform a maximum likelihood computation to improve the accuracy of the CCP estimation, also removing the dependency on the implementation of the linguistic model.
\item We establish a scale of CCP across projects, indicating that the metric provides information about the relative quality of different projects.
The scale shows that projects in the bottom decile of quality spend at least six times the effort on bug correction as projects in the top decile.
\item We show that CCP correlates with various other effects, e.g.\ successful onboarding of new developers and productivity.
\item We present twin experiments and co-change analysis in order to investigate relations beyond mere correlation.
\item On the way we also provide empirical support for Linus's law and the ``Quality is Free'' hypothesis.
\end{itemize}

\section{Related work}

Despite decades of work on this issue, there is no agreed definition of ``software quality''.
For some, this term refers to the quality of the software product as perceived by its users \cite{schneidewind02}.
Others use the term in reference to the code itself, as perceived by developers.
These two approaches have a certain overlap: bugs comprise bad code that has effects seen by the end user.
When considering the code, some define quality based on mainly non-functional properties, e.g.\ reliability, modifiability, etc.\ \cite{boehm76}.
Others include correctness as the foremost property \cite{dromey95}.
Our approach is also that correctness is the most important element of quality.
The number of bugs in a program could have been a great quality metric.
However, Rice's theorem \cite{10.2307/1990888} tells us that bug identification, like any non-trivial semantic property of programs, is undecidable.
Nevertheless, bugs are being found, providing the basis for the CCP metric.
And since bugs are time consuming, disrupt schedules, and hurt the general credibility, lowering the bug rate has value regardless of other implications---thereby lending value to having a low CCP.
Moreover, it is generally accepted that fixing bugs costs more the later they are found, and that maintenance is costlier than initial development \cite{boehm:econo, Boehm:2001:SDR:619059.621640,6191, Dawson2010SDLC}.
Therefore, the cost of low quality is even higher than implied by the bug ratio difference.

Capers Jones defined software quality as the combination of low defect rate and high user satisfaction \cite{Jones:1991:ASM:109758, CapersJonesQuality2012}.
He went on to provide extensive state-of-the-industry surveys based on defect rates and their correlation with various development practices, using a database of many thousands of industry projects.
Our work applies these concepts to the world of GitHub and open source, using the availability of the code to investigate its quality and possible causes and implications.

Software metrics can be divided into three groups: product metrics, code metrics, and process metrics.
Product metrics consider the software as a black box. 
A typical example is the ISO/IEC 25010:2011 standard \cite{ISO}. It includes metrics like fitness for purpose, satisfaction, freedom from risk, etc.\ 
These metrics might be subjective, hard to measure, and not applicable to white box actionable insights, which makes them less suitable for our research goals.
Indeed, studies of the ISO/IEC 9126 standard \cite{iec20019126} (the precursor of ISO/IEC 25010) found it to be ineffective in identifying design problems \cite{alkilidar05}.

Code metrics measure properties of the source code directly.
Typical metrics are lines of code (LOC) \cite{lipow1982number}, the Chidamber and Kemerer object oriented metrics (aka CK metrics) \cite{Chidamber:1994:MSO:630808.631131},
McCabe's cyclomatic complexity \cite{McCabe:1976:CM:1313324.1313586},
Halstead complexity measures \cite{Halstead:1977:ESS:540137},
etc.\ \cite{544352, 6606589, 1542070}.
They tend to be specific, low level and highly correlated with LOC \cite{shepperd88,rosenberg1997some, gil17, 10.1007/978-3-030-40223-5_8}.
Some specific bugs can be detected by matching patterns in the code \cite{Hovemeyer:2004:FBE:1052883.1052895}.
But this is not a general solution, since depending on it would bias our data towards these patterns.

Process metrics focus on the code's evolution.
The main data source is the source control system.
Typical metrics are the number of commits, the commit size, the number of contributors, etc.\ \cite{859533, 6606589,  Moser:2008:ARS:1414004.1414063}.
Process metrics have been claimed to be better predictors of defects than code metrics for reasons like showing where effort is being invested and having less stagnation \cite{Moser:2008:ARS:1414004.1414063, 6606589}.

Working with commits as the entities of interest is also popular in just in time (JIT) defect prediction \cite{6341763}.
Unlike JIT, we are interested in the probability and not in a specific commit being corrective.
We also focus on long periods, rather than comparing the versions before and after a bug fix, which probably reflects an improvement.
We examine work at a resolution of years, and show that CCP is stable, so projects that are prone to errors stay so, despite prior efforts to fix bugs.

Focusing on commits, we need a way to know if they are corrective.
If one has access to both a source control system and a ticket management system, one can link the commits to the tickets \cite{Bird:2009:FBB:1595696.1595716} and reduce the CCP computation to mere counting. 
Yet, the links between commits and tickets might be biased \cite{Bird:2009:FBB:1595696.1595716}. 
The ticket classification itself might have 30\% errors \cite{Herzig:2013:IBI:2486788.2486840}, and may not necessarily fit the researcher's desired taxonomy.
And integrating tickets with the code management system might require a lot of effort, making it infeasible when analysing thousands of projects.
Moreover, in a research setting the ticket management system might be unavailable, so one is forced to rely on only the source control system.

When labels are not available, one can use linguistic analysis of the commit messages as a replacement.
This is often done in defect prediction, where supervised learning can be used to derive models based on a labeled training set \cite{Ray:2014:LSS:2635868.2635922, 5463279,   Hall:2012:SLR:2420627.2420790}.

In principle, commit analysis models can be used to estimate the CCP, by creating a model and counting hits.
That could have worked if the model accuracy was perfect.
We take the model predictions and use the hit rate and the model confusion matrix to derive a maximum likelihood estimate of the CCP.
Without such an adaptation, the analysis might be invalid, and the hits of different models would have been incomparable.

Our work is also close to Software Reliability Growth Models (SRGM) \cite{544240, 859533, 1701965}.
In SRGM one tries to predict the number of future failures, based on bugs discovered so far, and assuming the code base is fixed.
The difference between us is that we are not aiming to predict future quality.
We identify current software quality in order to investigate the causes and implications of quality.

The number of bugs was used as a feature and indicator of quality before as absolute number \cite{Khomh:2012:FRI:2664446.2664475, Reddivari2019SoftwareQP}, per period \cite{Vasilescu:2015:QPO:2786805.2786850}, and per commit \cite{Shihab:2012:ISR:2393596.2393670, Amit:2019:RRB:3345629.3345631}.
We prefer the per commit version since it is agnostic to size and useful as a probability.

\section{Definition and Computation of the Corrective Commit Probability Metric}
\label{sect:CCP}

We now describe how we built the Corrective Commit Probability metric, in three steps:
\begin{enumerate}
  \item Constructing a \hyperref[sect:GoldStandard]{gold standard} data set of labeled commit samples, identifying those that are corrective (bug fixes).
  These are later used to learn about corrective commits and to evaluate the model.
  \item Building and evaluating a supervised learning linguistic \hyperref[sect:model]{model} to classify commits as either corrective or not.
  Applying the model to a project yields a hit rate for that project.
  \item Using \hyperref[sect:MLE]{maximum likelihood estimation} in order to find the most likely CCP given a certain hit rate.
\end{enumerate}

The need for the third step arises because the hit rate may be biased, which might falsify further analysis like using regression and hypothesis testing.
By working with the CCP maximum likelihood estimation we become independent of the model details and its hit rate.
We can then compare the results across projects, or even with results based on other researchers' models.
We can also identify outliers deviating from the common linguistic behavior (e.g., non-English projects), and remove them to prevent erroneous analysis.

Note that we are interested in the \emph{overall probability} that a commit is corrective.
This is different from defect prediction, where is the goal is to determine whether a specific commit is corrective.
Finding the probability is easier than making detailed predictions.
In analogy to coin tosses, we are interested only in establishing to what degree a coin is biased, rather than trying to predict a sequence of tosses.
Thus, if for example false positives and false negatives are balanced, the estimated probability will be accurate even if there are many wrong predictions.

\subsection{Building a Gold Standard Data Set}
\label{sect:GoldStandard}

The most straight forward way to compute the CCP is to use a change log system for the commits and a ticket system for the commit classification \cite{Bird:2009:FBB:1595696.1595716}, and compute the corrective ratio.
However, for many projects the ticket system is not available.
Therefore, we base the commit classification on linguistic analysis, which is built and evaluated using a gold standard.

A gold standard is a set of entities with labels that capture a given concept.
In our case, the entities are commits, the concept is corrective maintenance \cite{Swanson:1976:DM:800253.807723}, namely bug fixes, and the labels identify which commits are corrective.
Gold standards are used in machine learning for building models, which are functions that map entities to concepts.
By comparing the true label to the model's prediction, one can estimate the model performance.
In addition, we also used the gold standard in order to understand the data behavior and to identify upper bounds on performance.

We constructed the gold standard as follows.
Google's BigQuery has a \href{https://bigquery.cloud.google.com/dataset/bigquery-public-data:github_repos}{schema for GitHub} were all projects' commits are stored in a single table.
We sampled uniformly 840 (40 duplicate) commits as a train set.
The first author then manually labeled these commits as being corrective or not based on the commit content using a defined protocol.

To assess the subjectiveness in the labeling process, two additional annotators labeled 400 of the commits.
When there was no consensus, we checked if the reason was a deviation from the protocol or an error in the labeling (e.g., missing an important phrase).
In these cases, the annotator fixed the label. 
Otherwise, we considered the case as a disagreement. 
The final label was a majority vote of the annotators.
The Cohen's kappa scores \cite{Cohens1960Kappa} among the different annotators were at least 0.9, indicating excellent agreement.
Similarly consistent commit labeling was reported by Levin and Yehudai
\cite{Levin:2017:BAC:3127005.3127016}.

Of the 400 triple-annotated commits, there was consensus regarding the labels in 383 (95\%) of them: 105 (27\%) were corrective, 278 were not. 
There were only 17 cases of disagreement.
An example of disagreement is ``correct the name of the Pascal Stangs library.''
It is subjective whether a wrong name is a bug. 

In addition, we also noted the degree of certainty in the labeling.
The message ``mysql\_upgrade should look for .sql script also in share/ directory'' is clear, yet it is unclear whether the commit is a new feature or a bug fix.
In only 7 cases the annotators were uncertain and couldn't determine with high confidence the label from the commit message and content.
Of these, in 4 they all nevertheless selected the same label.

Two of the samples (0.5\%) were not in English. This prevents English linguistic models from producing a meaningful classification. 
Luckily, this is uncommon.

Finally, in 4 cases (1\%) the commit message did not contain any syntactic evidence for being corrective. The most amusing example was ``When I copy-adapted handle\_level\_irq I skipped note\_interrupt because I considered it unimportant.  
If I had understood its importance I would have saved myself some ours of debugging'' (the typo is in the origin).
Such cases set an upper bound on the performance of any syntactic model.
In our data set, all the above special cases (uncertainty, disagreement, and lack of syntactic evidence) are rather rare (just 22 samples, 5.5\%, many behaviors overlap), and the majority of samples are well behaved.
The number of samples in each misbehavior category is very small so ratios are very sensitive to noise. 
However, we can say with confidence that these behaviors are not common and therefore are not an issue of concern in the analysis.

\subsection{Syntactic Identification of a Corrective Commit}
\label{sect:model}

Our linguistic model is a supervised learning model, based on indicative terms that help identify corrective commit messages.
Such models are built empirically by analyzing corrective commit messages in distinction from other commit messages.

The most common approach today to do this is to employ machine learning.
We chose not to use machine learning classification algorithms to build the model. 
The main reason was that we are using a relatively small labeled data set, and linguistic analysis tends to lead to many features (e.g., in a bag of words, word embedding, or n-grams representation).
In such a scenario, models might overfit and be less robust. 
One might try to cope with overfitting by using models of low capacity. 
However, the concept that we would like to represent (e.g., include ``fix'' and ``error'' but not ``error code'' and ``not a bug'') is of relatively high capacity. 
The need to cover many independent textual indications and count them requires a large capacity, larger than what can be supported by our small labeled data set.
Note that though we didn't use classification algorithms, the goal, the structure, and the usage of the model are of supervised learning.

Many prior language models suggest term lists like (`bug', `bugfix',`error', `fail', `fix'), which reach 88\% accuracy on our data set.
We tried many machine learning classification algorithms and only the plain decision tree algorithm reached such accuracy.
More importantly, as presented later, we aren't optimizing for accuracy.
We therefore elected to construct the model manually based on several sources of candidate terms and the application of semantic understanding.

We began with a private project in which the commits could be associated to a ticket-handling system that enabled determining whether they were corrective. 
We used them in order to differentiate the word distribution of corrective commit messages and other messages and find an initial set of indicative terms. 
In addition, we used the gold-standard data-set presented above.
This data set is particularly important because our target is to analyze GitHub projects, so it is desirable that our train data will represent the data on which the model will run.
This train data set helped tuning the indicators by identifying new indications and nuances and alerting to bugs in the model implementation.

To further improving the model we used some terms suggested by Ray et al.\ \cite{Ray:2014:LSS:2635868.2635922}, tough we didn't adopt all of them (e.g., we don't consider a typo to be a bug).
This model was used in Amit and Feitelson \cite{Amit:2019:RRB:3345629.3345631}, reaching an accuracy of 89\%.
We then added additional terms from Shrikanth et al. \cite{shrikanth2019assessing}.
We also used labeled commits from Levin and Yehudai \cite{Levin:2017:BAC:3127005.3127016} to further improve the model based on samples it failed to classify.

The last boost to performance came from the use of active learning \cite{Settles10activelearning} and specifically the use of classifiers discrepancies \cite{archimedes}. 
Once the model's performance is high, the probability of finding a false negative, $positive\_rate \cdot (1-recall)$, is quite low, requiring a large number of manually labeled random samples per false negative.
Amit and Feitleson \cite{Amit:2019:RRB:3345629.3345631} provided models for a commit being corrective, perfective, or adaptive. 
A commit not labeled by any of the models is assured to be a false negative (of one of them).
Sampling from this distribution was an effective method to find false negatives, and improving the model to handle them increased the model recall from 69\% to 84\%.
Similarly, while a commit might be both corrective and adaptive, commits marked by more than one classifier are more likely to be false positives.

The resulting model uses regular expressions to identify the presence of different indicator terms in commit messages.
We base the model on straightforward regular expressions because this is the tool supported by Google's BigQuery relational database of \href{https://bigquery.cloud.google.com/dataset/bigquery-public-data:github_repos}{GitHub data}, which is our target platform.

The final model is based on three distinct regular expressions.
The first identifies about 50 terms that serve as indications of a bug fix.
Typical examples are: ``bug'', ``failure'', and ``correct this''.
The second identifies terms that indicate other fixes, which are not bug fixes.
Typical examples are: ``fixed indentation'' and ``error message''.
The third is terms indicating negation.
This is used in conjunction with the first regular expression to specifically handle cases in which the fix indication appears in a negative context, as in ``This is not an error''.
It is important to note that fix hits are also hits of the other fixes and the negation.
Therefore, the complete model counts the indications for a bug fix (matches to the first regular expression) and subtracts the indications for not really being a bug fix (matches to the other two regular expressions).
If the result is positive, the commit message was considered to be a bug fix.
The results of the model evaluation using a 1,100 samples test set built in Amit and Feitelson \cite{Amit:2019:RRB:3345629.3345631} are presented in the confusion matrix of Table \ref{tab:test-cm}.

\begin {table}[h!]\centering
\caption{ \label{tab:test-cm} Confusion matrix of model on test data set.}
\begin {tabular} { | l | l | l |}
\hline
                &\multicolumn{2}{c|}{Classification}              \\ \cline{2-3}
    Concept & True(Corrective) & False              \\ \hline
    True &  228 (20.73\%) TP &  43 (3.91 \%) FN      \\ \hline
    False &  34 (3.09\%) FP &  795 (72.27\%) TN \\ \hline
\end {tabular}
\end {table}

These results can be characterized by the following metrics:
\begin{itemize}
    \item Accuracy (model is correct): 93.0\%
    \item Precision (ratio of hits that are indeed positives): 87.0\%
    \item Precision lift ($\frac{precision}{positive\ rate}-1$): 253.2\%
    \item Hit rate (ratio of commits identified by model as corrective): 23.8\%
    \item Positive rate (real corrective commit rate): 24.6\%
    \item Recall (positives that were also hits): 84.1\%
    \item Fpr (False Positive Rate, negatives that are hits by mistake): 4.2\%
\end{itemize}

Though prior work was based on different protocols and data sets and therefore hard to compare, our accuracy is significantly better than previous reported results of 68\% \cite{5090025}, 70\% \cite{Amor_discriminatingdevelopment}, 76\% \cite{Levin:2017:BAC:3127005.3127016} and 82\% \cite{10.1145/1463788.1463819}, and also better than our own previous result of 89\% \cite{Amit:2019:RRB:3345629.3345631}.
The achieved accuracy is close to the well-behaving commits ratio in the gold standard.

\subsection{Maximum Likelihood Estimation of the Corrective Commit Probability}
\label{sect:MLE}

We now present the CCP maximum likelihood estimation.
Let $hr$ be the hit rate (probability that the model will identify a commit as corrective) and $pr$ be the positive rate, the true corrective rate in the commits (this is what CCP estimates).

In prior work it was common to use the hit rate directly as the estimate for the positive rate.
However, they differ since model prediction is not perfect.
Thus, by considering the model performance we can better estimate the positive rate given the hit rate.
From a performance modeling point of view, the Dawid-Skene \cite{10.2307/2346806} modeling is an ancestor of our work.
Though, the Dawid-Skene framework represents a model by its precision and recall, and we use Fpr and recall.

There are two distinct cases that can lead to a hit.
The first is a true positive (TP): There is indeed a bug fix and our model identifies it correctly. 
The probability of this case is $\Pr(TP) = pr \cdot recall$.
The second case is a false positive (FP): There was no bug fix, yet our model mistakenly identifies the commit as corrective. 
The probability of this case is $\Pr(FP) = (1-pr) \cdot Fpr$.
Adding them gives
\begin{equation}\label{eq:hr-from-pr}
hr = \Pr(TP) + \Pr(FP) = (recall - Fpr)pr + Fpr 
\end{equation}
Extracting $pr$ leads to
\begin{equation}
\label{eq:pr-from-hr}
pr = \frac{hr-Fpr}{recall-Fpr}
\end{equation}

We want to estimate $\Pr(pr|hr)$.
Let $n$ be the number of commits in our data set, and $k$ the number of hits.
As the number of samples increases, $\frac{k}{n}$ converges to the model hit rate $hr$. 
Therefore, we estimate $\Pr(pr|n, k)$. 
We will use maximum likelihood for the estimation. 
The idea behind maximum likelihood estimation is to find the value of $pr$ that maximizes the probability of getting a hit rate of $hr$.

Note that if we were given $p$, a single trial success probability, we could calculate the probability of getting $k$ hits out of $n$ trails using the binomial distribution formula 
\begin{equation}
\Pr(k;n,p) = \genfrac(){0pt}{0}{n}{k} \, p^k (1-p)^{n-k} 
\end{equation}

Finding the optimum requires the computation of the derivative and finding where it equals to zero.
The maximum of the binomial distribution is at $\frac{k}{n}$.
Equation (\ref{eq:pr-from-hr}) is linear and therefore monotone.
Therefore, the maximum likelihood estimation of the formula is
\begin{equation}\label{eq:mle_optimum} 
pr=\frac{\frac{k}{n}-Fpr}{recall-Fpr}
\end{equation}

For our model, $Fpr=0.042$ and $recall=0.84$ are fixed constants (rounded values taken from the confusion matrix of table \ref{tab:test-cm}).
Therefore, we can obtain the most likely $pr$ given $hr$ by 
\begin{equation}\label{eq:labeling_func_adaption} 
pr = \frac{hr-0.042}{0.84-0.042} =1.253\cdot hr -0.053
\end{equation}

\section{Validation of the CCP Metric}

\subsection{Validation of the CCP Maximum Likelihood Estimation}
\label{sect:MleValidation}

George Box said: ``All models are wrong but some are useful'' \cite{BOX1979201}.
We would like to see how close the maximum likelihood CCP estimations are to the actual results.
Note that the model performance results we presented above in Table \ref{tab:test-cm}, using the gold standard test set, do not refer to the maximum likelihood CCP estimation.
We need a new independent validation set to verify the maximum likelihood estimation.
We therefore manually labeled another set of 400 commits, and applying the model resulted in the confusion matrix shown in Table \ref{tab:MLE-validation-cm}.

\begin {table}[h!]\centering
\caption{ \label{tab:MLE-validation-cm} Confusion matrix of model on validation data set.}
\begin {tabular} { | l | l | l |}
\hline
                &\multicolumn{2}{c|}{Classification}              \\ \cline{2-3}
Concept & True(Corrective) & False              \\ \hline
True &  91 (22.75\%) TP &  18 (4.5\%) FN      \\ \hline
False &  34 (8.5\%) FP &  257 (64.25\%) TN \\ \hline
\end {tabular}
\end {table}

In this data set the positive rate is 27.2\%, the hit rate is 31.2\%, the recall is 83.5\%, and the Fpr is 11.7\%.
Note that the positive rate in the validation set is 2.6 percent points different from our test set.
The positive rate has nothing to do with MLE and shows that statistics tend to differ on different samples.
In this section we would like to show that the MLE method is robust to such changes.

\begin{figure}[h!]
\centering
\includegraphics[width=0.48\textwidth,trim={0mm 0mm 0 0mm},clip]{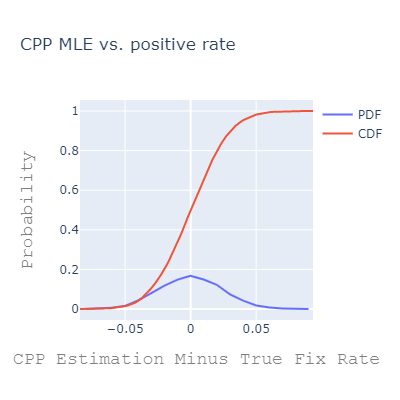}
\caption{\label{fig:mle-validation-dist}
Difference distribution in validation bootstrap.}
\end{figure}

In order to evaluate how sensitive the maximum likelihood estimation is to changes in the data, we used the bootstrap method \cite{Efron1992}.
We sampled with replacement 400 items from the validation set, repeating the process 10,000 times.
Each time we computed the true corrective commit rate, the estimated CCP, and their difference.
Figure \ref{fig:mle-validation-dist} shows the difference distribution.

In order to cover 95\% of the distribution we can trim the 2.5\% tails from both sides. 
This will leave us with differences ranging between -0.044 to 0.046. 
One can use the boundaries related to 95\%, 90\%, etc.\ in order to be extra cautious in the definition of the valid domain.

Another possible source of noise is in the model performance estimation.
If the model is very sensitive to the test data, a few anomalous samples can lead to a bad estimation.
Again, we used bootstrap in order to estimate the sensitivity of the model performance estimation.
For 10,000 times we sampled \emph{two} data sets of size 400.
On each of the data sets we computed the recall and Fpr and built an MLE estimator.
We then compared the difference in the model estimation at a few points of interest:
[0,1] -- the boundaries of probabilities, [0.042, 0.84] -- the boundaries of the valid domain, and [0.06, 0.39] -- the $p_{10}$ and $p_{90}$ percentiles of the CCP distribution.
Since our models are linear, so are their differences. 
Hence their maximum points are at the ends of the examined segments.
When considering the boundaries of probabilities [0,1], the maximal absolute difference is 0.34 and 95\% of the differences are lower than 0.19.
When considering the boundaries of the valid domain [0.042, 0.84], the maximal absolute difference is 0.28 and 95\% of the differences are lower than 0.15.
When considering the $p_{10}$ and $p_{90}$ percentiles of the CCP distribution [0.06, 0.39], the maximal absolute difference is 0.13 and 95\% of the differences are lower than 0.07.

Using the validation set estimator on the test set, the CCP is 0.168, 7.7 percentage points off the actual positive rate.
In the other direction, using the CCP estimator test data performance on the validation set, the CCP is 0.39, 11.8 points off.
Since our classifier has high accuracy, the difference between the hit rate and the CCP estimates in the distribution deciles, presented below in Table \ref{tab:CCP-distrib}, is at most 4 percentage points. 
Hence the main practical contribution of the MLE in this specific case is the \hyperref[sect:fixec-linguistic-model]{identification of the valid domain} rather than an improvement in the estimate.

\subsection{Sensitivity to Violated Assumptions}

\subsubsection{Fixed Linguistic Model Assumption}
\label{sect:fixec-linguistic-model}

The maximum likelihood estimation of the CCP assumes that the linguistic model performance, measured by its $recall$ and $Fpr$, is fixed.
Hence, a change in the hit rate in a given domain is due to a change in the CCP in the domain, and not due to a change in the linguistic model performance.
This assumption is crucial for the mapping from identified corrective commits to a quality metric.

Yet, this assumption does not always hold. 
Both $hr$ and $pr$ are probabilities and must be in the range $[0,1]$.
Equation (\ref{eq:pr-from-hr}) equals 0 at $Fpr$ and 1 at $recall$.
For our model, this indicates that the range of values of $hr$ for which $pr$ will be a probability is $[0.042,0.84]$. 
Beyond this range, we are assured that the linguistic model performance is not as measured on the gold standard. 
An illustrative example of the necessity of the range is a model with $recall=0.5$ and $Fpr=0$. 
Given $hr=0.9$ the most likely $pr$ is $1.8$. 
This is an impossible value for a probability, so we deduce that our assumption is wrong.

As described in Section \ref{sect:results}, we estimated the CCP of all 8,588 large active projects in 2019.
In 10 of these projects the estimated CCP was above 1.
Checking these projects, we found that they have many false positives, e.g.\ due to a convention of using the term ``bug'' for general tasks or starting the subject with ``fixes \#123'' where ticket \#123 was not a bug fix but some other task id.

Another 11.8\% of the projects had an estimated CCP below 0.
This could indicate having extremely few bugs, or else a relatively high fraction of false negatives (bug fixes we did not identify).
One possible reason for low identification is if the project commit messages are not in English.
To check this, we built a simple linguistic model in order to identify if a commit message is in English.
The model was the 100 most frequent words in English longer than two letters (see details and performance in supplementary materials). 
The projects with negative CCP had a median English hit rate 0.16.
For comparison, the median English hit rate of the projects with positive CCP was 0.54, and 96\% of them had a hit rate above 0.16.

Interestingly, another reason for many false negatives was the habit of using very terse messages.
We sampled 5,000 commits from the negative CCP projects and compared them to the triple-annotated data set used above.
In the negative CCP commits, the median message length was only 27 characters, and the 90th percentile was 81 characters.
In the annotated data set the median was 8 times longer, and the 90th percentile was 9 times longer.

It is also known that not all projects in GitHub (called there repositories) are software projects \cite{Munaiah17Curating, Kalliamvakou15Promises}.
Since bugs are a software concept, other projects are unlikely to have such commits and their CCP will be negative.
Hence, the filtering also helps us to focus on software projects.
Git is unable to identify the language of 6\% of the projects with negative CCP, more than 14 time the ratio in the valid domain.
The languages `HTML', `TeX', `TSQL', `Makefile', `Vim script', `Rich Text Format' and `CSS' are identified for 22\% of the projects with negative CCP, more than 4 times as in the valid range.
Many projects involve some languages and when we examined a sample of projects we found that the language identification is not perfect.
However, at least 28\% of the projects that we filtered due to negative CCP are not identified by GitHub as regular software projects.

To summarize, in the projects with invalid CCP estimates, below 0 or above 1, the behavior of the linguistic model changes and invalidates the fixed performance assumption.
We believe that the analysis of projects in the CCP valid domain is suitable for software engineering goals.
The CCP distribution in Table \ref{tab:CCP-distrib} below is presented for both the entire data set and only for projects with valid CCP estimates.
The rest of the analysis is done only on the valid projects.

\subsubsection{Fixed Bug Detection Efficiency Assumption}
\label{sect:DetectionEfficiency}

The major assumption underlying our work is that CCP reflects quality --- that the number of bug fixes reflects the number of bugs.
Likewise, the comparison of CCP across projects assumes that the bug detection efficiency is similar, so a difference in the CCP is due to a difference in the existence of bugs and not due to a difference in the ability to find them.
We found two situations in which this assumption appears to be violated.
In these situations, the ability to find bugs appears to be systematically different --- higher or lower --- than in other projects.

The first such situation is in very popular projects.
\emph{Linus's law}, ``given enough eyeballs, all bugs are shallow'' \cite{LinusRule}, suggests that a large community might lead to more effective bug identification, and as a consequence also to higher CCP.
In order to investigate this, we used projects of companies or communities known for their high standards: Google, Facebook, Apache, Angular, Kubernetes, and Tensorflow.
For each such source, we compared the average CCP of projects in the top 5\% as measured by stars (7,481 stars or more), with the average CCP of projects with fewer stars.

\begin{table}[h!]\centering
\caption{\label{tab:CCP-linus}
Linus's Law: CCP in projects with many or fewer stars.}
\begin {tabular}{ | l | c | c | c | c |}
\hline
& \multicolumn {2} {c |} {top 5\%} & \multicolumn {2} {c |} {bottom 95\%}\\ 
& \multicolumn {2} {c |} {(>7,481 stars)} & \multicolumn {2} {c |} {(<7,481 stars)}\\ \cline {2 - 5}
Source     & $N$ & avg.\ CCP (lift) & $N$ & avg.\ CCP \\ \hline
Google     &  8  & 0.32 (27\%) &  66   & 0.25 \\ \hline
Facebook   &  9  &  0.30 (12\%) &  9 & 0.27 \\ \hline
Apache     &  10  &  0.37 (44\%) &  35 &  0.26 \\ \hline
Angular    &  3  &  0.49 (34\%) &  32 &  0.37 \\ \hline
Kubernetes & 3   &  0.21 (35\%)  & 3 &  0.16 \\ \hline
Tensorflow &  5  & 0.26 (32\%) &  26 &  0.20\\ \hline
\end{tabular}
\end{table}

The results were that the most popular projects of high-reputation sources indeed have CCP higher than less popular projects of the same organization (Table \ref{tab:CCP-linus}).
The popular projects tend to be important projects: Google's Tensorflow and Facebook's React received more than 100,000 stars each.
It is not likely that such projects have lower quality than the organization's standard.
Apparently, these projects attract large communities which provide the eyeballs to identify the bugs efficiently, as predicted by Linus's law.

Note that these communities' projects, with many stars or not, have average CCP of 0.26, 21\% more than all projects' average.
Their average number of authors is 219, 144\% more than the others.
Their average age is 4 years compared to 5 years, 20\% younger yet not young in both cases.
However, the average number of stars is 5,208 compared to 1,428, a lift of 364\%.
It is possible that while the analysis we presented is for extreme numbers of stars, Linus's law kicks in already at much lower numbers and contributed to the difference.

There are only few such projects (we looked at the top 5\% from a small select set of sources).
The effect on the CCP is modest (raising the level of bug corrections by around 30\%, both a top and a median project will decrease in one decile).
Thus, we expect that they will not have a significant impact on the results presented below.

The mirror image of projects that have enough users to benefit from Linus's law is projects that lose their core developers.
The ``Truck Factor'' originated in the Agile community.
Its informal definition is ``The number of people on your team who have to be hit with a truck before the project is in serious trouble'' \cite{10.5555/548833}.
In order to analyze it, we used the metric suggested by Avelino et al. \cite{DBLP:journals/corr/AvelinoPHV16}.
Truck Factor Developers Detachment (TFDD) is the event in which the core developers abandon a project as if a virtual truck had hit them \cite{DBLP:journals/corr/abs-1906-08058}.
We used instances of TFDD identified by Avelino et al.\ and matched them with the GitHub behavior \cite{DBLP:journals/corr/abs-1906-08058}.
As expected, TFDD is a traumatic event for projects, and 59\% of them do not survive it.

When comparing 1-month windows around a TFDD, the average number of commits is reduced by 1 percentage point.
There is also an average reduction of 3 percentage points in refactoring, implying a small decrease in quality improvement effort.
At the same time, the CCP improves (decreases) by 5 percentage points.
Assuming that quality is not improved as a result of a TFDD, a more reasonable explanation is that bug detection efficiency was reduced.
But even the traumatic loss of the core developers damage is only 5 percentage points.

The above cases happen in identifiable conditions, and therefore could be filtered out.
But since they happen in extreme, rare cases, we choose to leave them, and gain an analysis that though slightly biased, represents the general behaviour of projects.

\subsection{CCP as a Quality Metric}

The most important property of a metric is obviously its validity, that it reflects the measured concept.
There is a challenge in showing that CCP measures quality since there is no agreed definition of quality.

\begin{figure}[]
\centering
\includegraphics[scale=0.4]{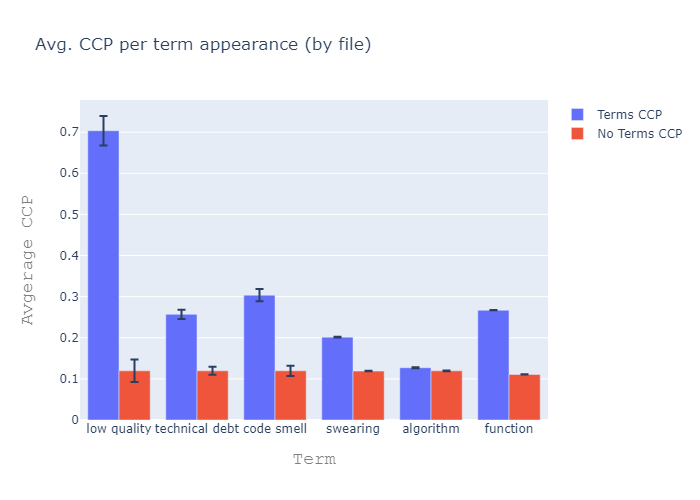}
\caption{\label{fig:ccp_by_quality_terms}
CCP of files with or without different quality terms.
}
\end{figure}
\begin{figure}[]
\centering
\includegraphics[scale=0.4]{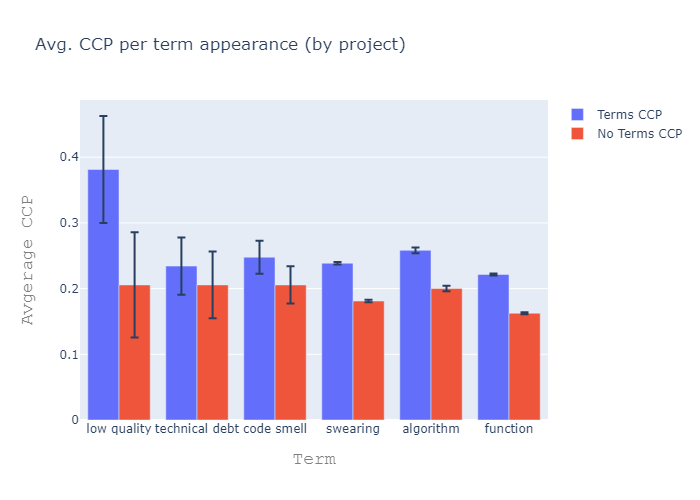}
\caption{\label{fig:ccp_by_quality_terms_by_repo}
CCP of projects with or without different quality terms.
}
\end{figure}

To circumvent this, we checked references to low quality in commit messages and correlated them with CCP (Figs.\ \ref{fig:ccp_by_quality_terms} and \ref{fig:ccp_by_quality_terms_by_repo}).
The specific terms checked were direct references to ``low quality'', and related terms like ``code smell'' \cite{fowler1997refactoring, van2002java, yamashita2012code, khomh2009exploratory, 6676898}, and ``technical debt'' \cite{10.1145/157709.157715, tom2013exploration, kruchten2012technical}.
In addition, swearing is also a common way to express dissatisfaction, with millions of occurrences compared to only hundreds or thousands for the technical terms.
For files, we considered files with 10+ commits and compared those with at least 10\% occurrences of the term to the rest.
Projects may have a lot of commits, and the vast majority do not contain the terms.
Instead of a ratio, we therefore consider a project to contain a term if it has at least 10 occurrences.
As the figures show, when the terms appear, the CCP is higher (sometime many times higher).
Thus, our quality metric agrees with the opinions of the projects' developers regarding quality.

To verify this result, we attempted to use negative controls.
A negative control is an item that should be indifferent to the analysis.
In our case, it should be a term not related to quality.
We choose ``algorithm'' and ``function'' as such terms.
The verification worked for ``algorithm'' at the file level: files with and without this term had practically the same CCP.
But files with ``function'' had a much higher CCP than files without it, and projects with both terms had a higher CCP that without them.
Possible reasons are some relation to quality (e.g., algorithmic oriented projects are harder) or biases (e.g., object-oriented languages tend to use the term ``method'' rather than ``function'').
Anyway, it is clear that the difference in ``low quality'' is much larger and there is a large difference in the other terms too.
Note that this investigation is not completely independent. 
While the quality terms used here are different from those used for the classification of corrective commits, we still use the same data source.

\hide{
It seems that it is quite hard to find a pure negative control and you can see that the same terms behave differently at the file and the project resolutions.

}%

An additional important attribute of metrics is that they be stable.
We estimate stability by comparing the CCP of the same project in adjacent years, from 2014 to 2019.
Overall, the quality of the projects is stable over time. 
The Pearson correlation between the CCP of the same project in two successive years, with 200 or more commits in each, is 0.86.
The average CCP, using all commits from all projects, was 22.7\% in 2018 and 22.3\% in 2019.
Looking at projects, the CPP grew on average by 0.6 percentage points from year to year, which might reflect a slow decrease in quality.
This average hide both increases and decreases; the average absolute difference in CPP was 5.5 percentage points.
Compared to the CCP distribution presented in Table \ref{tab:CCP-distrib} below, the per project change is very small.

\section{Association of CCP with Project Attributes}
\label{sect:CcpProjAssoc}

To further support the claim that CCP is related to quality, we studied the correlations of CCP with various notions of quality reflected in project attributes.
To strengthen the results beyond mere correlations we control for variables which might influence the results, such as project age and the number of developers.
We also use \hyperref[sect:CcpCoChange]{co-change analysis} and \hyperref[sect:CcpTwin]{``twin'' analysis}, which show that the correlations are consistent and unlikely to be random.

\subsection{Methodology}
\label{sect:CcpMethod}

Our results are in the form of correlations between CCP and other metrics.
For example, we show that projects with shorter files tend to have a lower CCP.
These correlations are informative and actionable, e.g., enabling a developer to focus on longer files during testing and refactoring.
But correlation is not causation, so we cannot say conclusively that longer files \emph{cause} a higher propensity for bugs that need to be fixed.
Showing causality requires experiments in which we perform the change, which we leave for future work.
The correlations that we find indicate that a search for causality might be fruitful and could motivate changes in development practices that may lead to improved software quality.

In order to make the results stronger than mere correlation, we use several methods in the analysis.
We \hyperref[sect:CcpControl]{control the results} to see that the relation between A and B is not due to C.
In particular we \hyperref[sect:CcpTwin]{control for the developer}, by observing the behaviour of the same developer in different projects.
This allows us to separate the influence of the developer and the project.
We use \hyperref[sect:CcpCoChange]{co-change over time analysis} in order to see to what extent a change in one metric is related to a change in the other metric.

The distributions we examined tended to have some outliers that are much higher than the mean and the majority of the samples.
Including outliers in the analysis might distort the results.
In order to reduce the destabilizing effect of outliers, we applied Winsorizing \cite{hastings1947}.
We used one-sided Winsorizing, where all values above a certain threshold are set to this threshold.
We do this for the top 1\% of the results throughout, to avoid the need to identify outliers and define a rule for adjusting the threshold for each specific case.
In the rest of the paper we used the term capping (a common synonym) for this action.
In addition, we check whether the metrics are stable across years.
A reliable metric applied to clean data is expected to provide similar results in successive years.

Results are computed on 2019 active projects, and specifically on projects whose CCP is in the valid domain.
We didn't work with version releases since we work with thousands of projects whose releases are not clearly marked.
Note that in projects doing continuous development, the concept of release is no longer applicable.

\subsubsection{Controlled Variables: Project Age, Number of Developers, and Programming Language}
\label{sect:CcpControl}

Our goal is to find out how to improve software development.
We would like to provide actionable recommendations for better software development.
However, there are factors that influence software quality that are hard to change.
It is not that helpful for an ongoing project to find out that a different programming language is indicative of a lower bug rate.
Yet, we examine the effect of some variables that influence the quality yet are hard to change.
We control them in the rest of the analysis to validate that the results hold.
We do the control by conditioning on the relevant variable and checking if the relations found in general hold while controlling too.
We don't control by more then one variable at a time since our data set is rather small and controlling leads to smaller data sets, making the results less robust to noise. 

\begin{figure}[h!]
\centering
\includegraphics[scale=0.35]{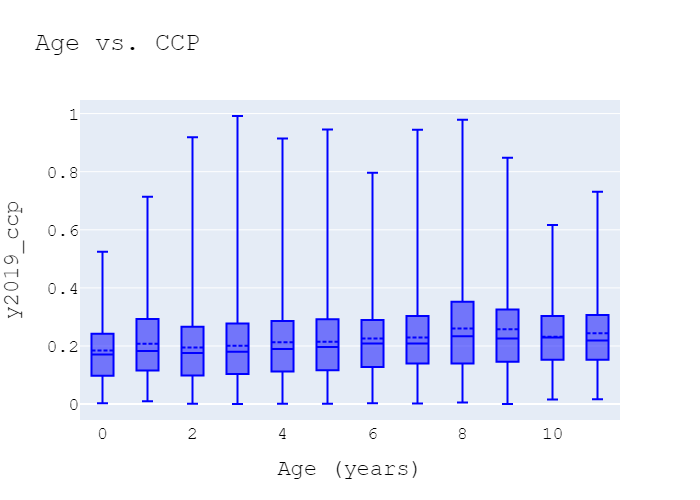}
\caption{\label{fig:age}
CCP distribution (during 2019) in projects of different ages.
In this and following figures, each boxplot shows the 5, 25, 50, 75, and 95 percentiles. The dashed line represents the mean.
}
\end{figure}

Lehman's laws of software evolution imply that quality may have a negative correlation with the age of a project \cite{1456074, lehman97}.
We checked this on our dataset.
We first filtered out projects that started before 2008 (GitHub beginning).
For the remaining projects, we checked their CCP at each year.
Figure \ref{fig:age} shows that CCP indeed tends to increase slightly with age.
In the first year, the average CCP is 0.18.
There is then a generally upward trend, getting to an average of 0.23 in 10 years.
Note that there is a survival bias in the data presented since many projects do not reach high age.

In order to see that our results are not due to the influence of age, we divided the projects into age groups.
Those started earlier than 2008 were excluded, those started in 2018--2019 (23\%) are considered to be young, the next, from 2016--2017 (40\%), are medium, and those from 2008--2015 (37\%) are old.
When we obtained a result (e.g., correlation between coupling and CCP), we checked if the result holds for each of the groups separately.

The number of developers, via some influence mechanisms (e.g., ownership), was investigated as a quality factor and it seems that there is some relation to quality \cite{norick2010effects, bird2011don, Weyuker08Spoil}.
The number of developers and CCP have Pearson correlation of 0.12.
The number of developers can reach very high values and therefore be very influential.

\begin{figure}[h!]
\centering
\includegraphics[scale=0.35]{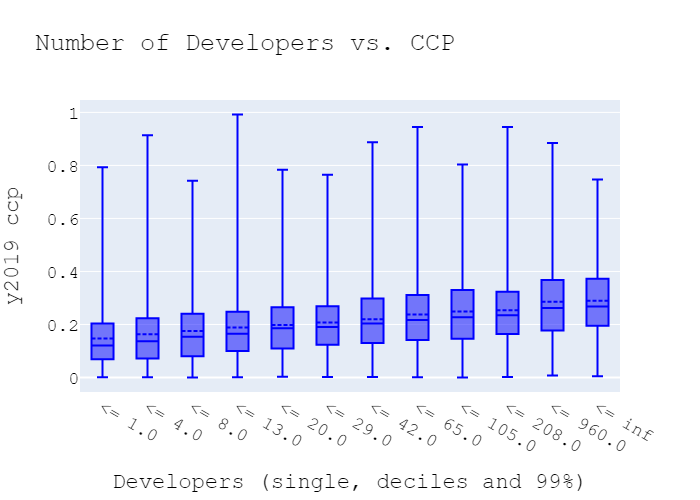}
\caption{\label{fig:authors_ccp}
CCP distribution for projects with different numbers of developers.
}
\end{figure}

Fig.\ \ref{fig:authors_ccp} shows that percentiles of the CCP distribution increase monotonically with the number of developers.
Many explanations have been given to the quality reduction as the number of developers increases.
It might be simply a proxy to the project size (i.e.\ to the LOC).
It might be to the increased communication complexity and the difficulty to coordinate multiple developers, as suggested by Brooks in the mythical ``The Mythical Man Month'' \cite{brooks:book}.
Part of it might also be a reflection of Linus's law, as discussed in Section \ref{sect:DetectionEfficiency}.

We control for the number of developers by grouping the 25\% of project with the least developers as few (at most 10), the next 50\% as intermediate (at most 80), and the rest as numerous, and verifying that results hold for each such group.

Results regarding the influence of programming language are presented below in \hyperref[sect:CcpLang]{Section \ref{sect:CcpLang}}.
We show that the projects written in different programming languages exhibit somewhat different distributions of CCP.
We therefore control for the programming language in order to see that our results remain valid for each language individually.

\subsubsection{Co-change Over Time}
\label{sect:CcpCoChange}

While experiments can help to determine causality, they are based on few cases and expensive.
On the other hand, we have access to plenty of observations, in which we can identify correlations.
While casual relations tend to lead to correlation, non-casual relations might also lead to correlations due to various reasons.
We would like to use an analysis that will help to filter out non-casual relations.
By that we will be left with a smaller set of more likely relations to be further investigated for causality.

When two metrics change simultaneously, it is less likely to be accidental.
Hence, we track the metrics over time in order to see how their changes match.
We create pairs of the same project in two consecutive years.
For each pair we mark if the first and second metrics improved.
We observe the following co-changes.
The ratio of improvement match (the equivalent to accuracy in supervised learning), is an indication of related changes.
\newcommand{\imp}[1]{\mbox{$m_{#1}\!\!\uparrow$}}
Denote the event that metric $i$ improved from one year to the next by \imp{i}.
The probability $P(\imp{j} \: | \: \imp{i})$, (the equivalent to precision in supervised learning), indicates how likely we are to observe an improvement in metric $j$ knowing of an improvement in metric $i$.
It might be that we will observe high precision but it will be simply since $P(\imp{j})$ is high.
In order to exclude this possibility, we also observe the precision lift,
$\frac{P(\imp{j} \: | \: \imp{i})}{P(\imp{j})} -1$.
Note that lift cannot be used to identify the causality direction since it is symmetric:
\begin{equation}\label{eq:precision-lift}
\frac{P(\imp{j} \: | \: \imp{i})}{P(\imp{j})}= \frac{P(\imp{i} \: \wedge \: \imp{j})}{P(\imp{i}) \cdot P(\imp{j})}=\frac{P(\imp{i} \: | \: \imp{j})}{P(\imp{i})}
\end{equation}
If an improvement in metric $i$ indeed causes the improvement in metric $j$, we expect high precision and lift.
Since small changes might be accidental, we also investigate improvements above a certain threshold.
There is a trade-off here since given a high threshold the improvement is clear yet the number of cases we consider is smaller.
Another trade-off comes from how far in the past we track the co-changes.
The earlier we will go the more data we will have.
On the other hand, this will increase the weight of old projects, and might subject the analysis to changes in software development practices over time and to data quality problems.
We chose a scope of 5 years, avoiding looking before 2014.

\subsubsection{Controlling the Developer}
\label{sect:CcpTwin}

Measured metric results (e.g., development speed, low coupling) might be due to the developers working on the project (e.g., skill, motivation) or due to the project environment (e.g., processes, technical debt).
To separate the influence of developers and environment,
we checked the performance of developers active in more than one project in our data set.
By fixing a single developer and comparing the developer's activity in different projects, we can investigate the influence of the project.
Note that a developer active in $n$ projects will generate $O(n^2)$ project pairs (``twins'') to compare.

We considered only involved developers, committing at least 12 times per year, otherwise the results might be misleading.
While this omits 62\% of the developers, they are responsible for only 6\% of the commits.

Consider development speed as an example.
If high speed is due to the project environment, in high speed projects every developer is expected to be faster than himself in other projects.
This control resembles twin experiments, popular in psychology, where a behavior of interest is observed on twins. 
Since twins have a very close genetic background, a difference in their behavior is more likely to be due to another factor (e.g., being raised in different families).

Assume that performance on project A is in general better than on project B.
We consider developers that contributed to both projects, and check how often they are better in project A than themselves in project B
(formally, the probability that a developer is better in project A than in project B given that project A is better than project B).
This is equivalent to precision in supervised learning, where the project improvement is the classifier and the developer improvement is the concept.
In some cases, a small difference might be accidental.
Therefore we require a large difference between the projects and between the developer performance (e.g., at least 10 commits per year difference, or more formally, the probability that a developer committed at least 10 times more in project A than in project B  given that the average number of commits per developer in project A is at least 10 commits higher than in project B).

\subsubsection{Selection of Projects}

In 2018 Github published that they had \href{https://github.blog/2018-11-08-100m-repos/}{100 million projects}.  
The BigQuery GitHub schema contains about 2.5 million \emph{public} projects prior to 2020.
But the vast majority are not appropriate for studies of software engineering, being small, non-recent, or not even code.

\begin{figure}[h!]
\centering
\includegraphics[width=0.35\textwidth]{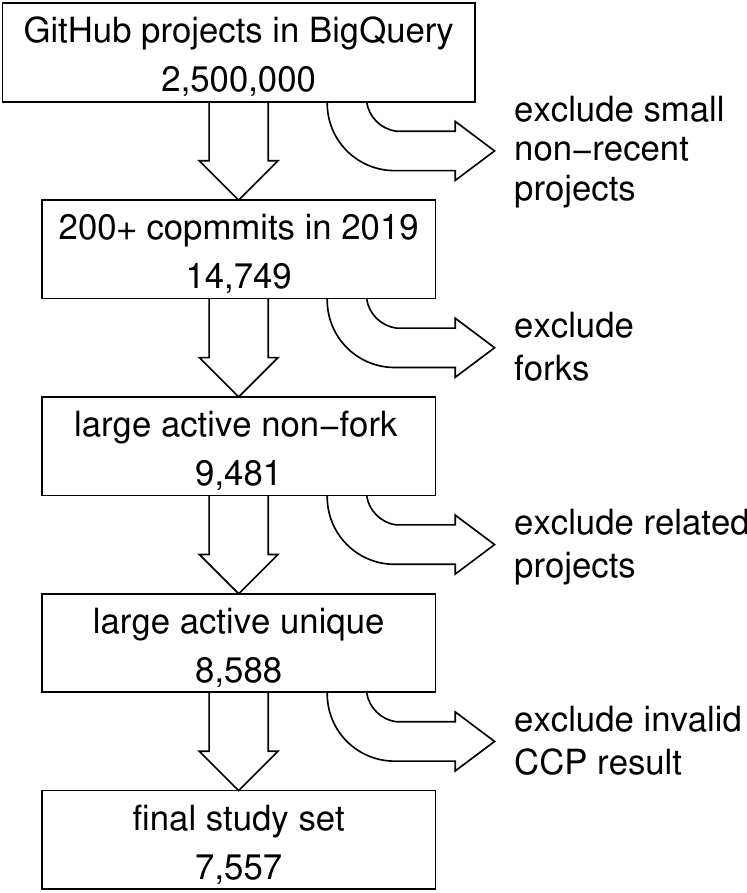}
\caption{\label{fig:proj-select}
Process for selecting projects for analysis.}
\end{figure}

In order to omit inactive or small projects where estimation might be noisy, we defined our scope to be all open source projects included in GitHub's BigQuery data with 200+ commits in 2019.
We selected a threshold of 200 to have enough data per project, yet have enough projects above the threshold.
There are 14,749 such projects (Fig.\ \ref{fig:proj-select}).

However, this set is redundant in the sense that some projects are closely related \cite{Kalliamvakou15Promises}.
The first step to reduce redundancy is to exclude projects marked in the GitHub API as being forks of other projects.
This reduced the number to 9,481 projects.
Sometimes extensive amounts of code are cloned without actual forking.
Such code cloning is prevalent and might impact analysis \cite{Gharehyazie2019, 10.1145/3133908, 10.1145/3359591.3359735}.
Using commits to identify relationships \cite{mockus2020complete}, we excluded dominated projects, defined to have more than 50 common commits with another, larger project, in 2019.
Last, we identified projects sharing the same name (e.g., `spark') and preferred those that belonged to the user with more projects (e.g., `apache').
After the redundant projects removal, we were left with 8,588 projects.
But calculating the CCP on some of these led to invalid values as described above.
For analysis purposes we therefore consider only projects where CCP is in the valid range, whose number is 7,557.

\subsection{Results}
\label{sect:results}

The following examples aim to show the applicability of CCP.
We compute the CCP of many projects and produce the \hyperref[sect:CcpDist]{CCP distribution}. 
We then demonstrate associations between high quality, as represented by CCP, and short \hyperref[sect:CcpLength]{source code file length}, \hyperref[sect:CcpCoupling]{coupling}, and \hyperref[sect:CcpLang]{programming language}.
We also investigate possible implications like \hyperref[sect:CcpDeveloper Engagement]{developer engagment} and \hyperref[sect:CcpSpeed]{development speed}.

\subsubsection{The Distribution of CCP per Project}
\label{sect:CcpDist}

Given the ability to identify corrective commits, we can classify the commits of each project and estimate the distribution of CCP over the projects' population.

\begin{table}[h!]\centering
\caption{\label{tab:CCP-distrib}
CCP distribution in active GitHub projects.}
\begin {tabular}{ | c | c | c | c | c |}
\hline
& \multicolumn {2} {c |} {Full data set} & \multicolumn {2} {c |} {CCP $\in [0, 1]$}\\
& \multicolumn {2} {c |} {( 8,588 projects)} &\multicolumn
{2}{c |}{( 7,557 projects)}\\ \cline {2 - 5}
Percentile &  Hit rate & CCP est. & Hit rate & CCP est.  \\ \hline
10 &  0.34  & 0.38  & 0.35  & 0.39 \\ \hline
20 &  0.28  & 0.30  & 0.29  & 0.32 \\ \hline
30 &  0.24  & 0.25  & 0.26  & 0.27 \\ \hline
40 &  0.21  & 0.21  & 0.22  & 0.23 \\ \hline
50 &  0.18  & 0.18  & 0.20  & 0.20 \\ \hline
60 &  0.15  & 0.14  & 0.17  & 0.17 \\ \hline
70 &  0.12  & 0.10  & 0.15  & 0.13 \\ \hline
80 &  0.09  & 0.06  & 0.12  & 0.10 \\ \hline
90 &  0.03  & -0.02  & 0.09  & 0.06 \\ \hline
95 &  0.00  & -0.05  & 0.07  & 0.04 \\ \hline
\end{tabular}
\end{table}

Table \ref{tab:CCP-distrib} shows the distribution of hit rates and CCP estimates on the GitHub projects with 200+ commits in 2019, with redundant repositories (representing the same project) excluded.
The hit rate represents the fraction of commits identified as corrective by the linguistic model, and the CCP is the maximum likelihood estimation.
The top 10\% of projects have a CCP of up to 0.06.
The median project has a CCP of 0.2, more than three times the top projects' CCP.
Interestingly, Lientz at el.\ reported a median of 0.17 in 1978, based on a survey of 69 projects \cite{lientz78}.
The bottom 10\% have a CCP of 0.39 or more, more than 6 times higher than the top 10\%.

Given the distribution of CCP, any developer can find the placement of his own project relative to the whole community.
The classification of commits can be obtained by linking them to tickets in the ticket-handling system (such as Jira or Bugzilla).
For projects in which there is a single commit per ticket, or close to that, one can compute the CCP in the ticket-handling system directly, by calculating the ratio of bug tickets.
Hence, having full access to a project, one can compute the exact CCP, rather than its maximum likelihood estimation.

Comparing the project's CCP to the distribution in the last column of Table \ref{tab:CCP-distrib} provides an indication of the project's code quality and division of effort calibrated with respect to other projects.

\subsubsection{File Length and CCP}
\label{sect:CcpLength}

The correlation between high file length and an increase in bugs has been widely investigated and considered to be a fundamental influencing metric \cite{lipow1982number, gil17}.
The following analysis first averages across files in each project, and then considers the distribution across projects, so as to avoid giving extra weight to large projects.
In order to avoid sensitivity due to large values, we capped large file lengths at 181KB, the 99th percentile.

\begin{figure}[h!]
\centering
\includegraphics[scale=0.35,trim={0mm 0mm 0 30mm},clip]{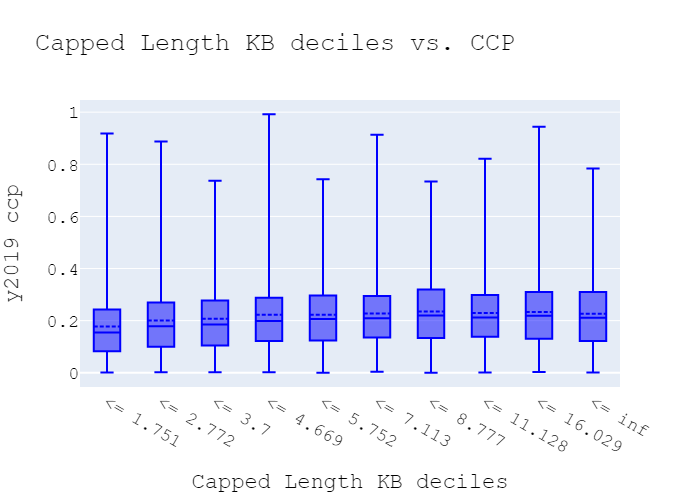}
\caption{\label{fig:ccp_by_Capped_Length_KB}
CCP distribution for files with different lengths (in KB, capped).
}
\end{figure}

In our projects data set, the mean file length was 8.1 KB with a standard deviation of 14.3KB, a ratio of 1.75 (capped values).
Figure \ref{fig:ccp_by_Capped_Length_KB} shows that the CCP increases with the length.
Projects whose average capped file size is in the lower 25\% (bellow 3.2KB) has average CCP of 0.19.
The last five deciles all have CCP around 0.23 as if at a certain point a file is ``just too long''.

We did not perform a co-change analysis of file length and CCP since the GitHub BigQuery database stores only the content of the files in the HEAD (last version), and not previous ones.

Controlling by project age and developers support the results.
When controlling for language, in most languages the top-ranked projects indeed have shorter files.
On the other hand, in PHP they are 10\% longer, and in JavaScript the lengths in the top 10\% quality projects is 31\% higher than the rest.

\subsubsection{Coupling and CCP}
\label{sect:CcpCoupling}

A commit is a unit of work ideally reflecting the completion of a task.
It should contain only the files relevant to that task.
Many files needed for a task means coupling.
Therefore, the average number of files in a commit can be used as a metric for coupling \cite{1231213, Amit:2019:RRB:3345629.3345631}.

To validate that this metric captures the way developers think about coupling, we compared it to the appearance of the terms ``coupled'' or ``coupling'' in messages of commits containing the file.
Out of the files with at least 10 commits, those with a hit rate of at least 0.1 for these terms had average commit size 45\% larger than the others.

When looking at the size of commits, it turns out that corrective commits involve significantly fewer files than other commit types:
the average corrective commit size is 3.8, while the average non-corrective commit size is 5.5.
Therefore, comparing files with different ratios of corrective commits will influence the apparent coupling.
To avoid this, we will compute the coupling using only non-corrective commits.
We define the coupling of a project to be the average coupling of its files.

\begin{figure}[h!]
\centering
\includegraphics[scale=0.35]{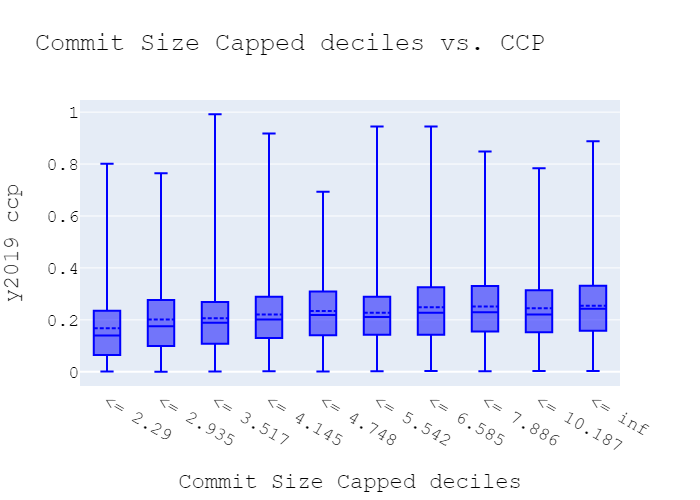}
\caption{\label{fig:ccp_by_coupling}
CCP distribution for projects with different average commit sizes (number of files, capped, in non-corrective commits).
}
\end{figure}

Figure \ref{fig:ccp_by_coupling} presents the results.
There is a large difference in the commit sizes:
the 25\% quantile is 3.1 files and the 75\% quantile is 7.1.
Similarly to the relation of CCP to file sizes, here too the distribution of CCP in commits above the median size appears to be largely the same, with an average of 0.24.
But in smaller commits there is a pronounced correlation between CCP and commit size, and the average CCP in the low coupling 25\% is 0.18.
Projects that are in the lower 25\% in both file length and coupling have 0.15 average CCP and 29.3\% chance to be in the top 10\% of the CCP-based quality scale, 3 times more than expected.

When we analyze CCP and coupling co-change, the match for any improvement is 52\%.
A 10-percentage point reduction in CCP and a one file reduction in coupling are matched 72\% of the time.
Given a reduction of coupling by one file, the probability of a CCP reduction of 10 percentage points is 9\%, a lift of 32\%.
Results hold when controlling for language, number of developers, and age, though in some setting the groups are empty or very small.

In twin experiments, the probability that the developer's coupling is better (lower) in the better project was 49\%, a lift of 15\%.
When the coupling in the better project was better by at least one file, the developer coupling was better by one file in 33\% of the cases, a lift of 72\%.

\subsubsection{Programming Languages and CCP}
\label{sect:CcpLang}

Our investigation of programming languages aims to control the influence on CCP, not to investigate programming languages as a subject.
Other than the direct language influence, languages are often used in different domains, and indirectly imply programming culture and communities.

\begin{figure}[h!]
\centering
 \includegraphics[width=0.7\textwidth,trim={0mm 0mm 0mm 10mm},clip]{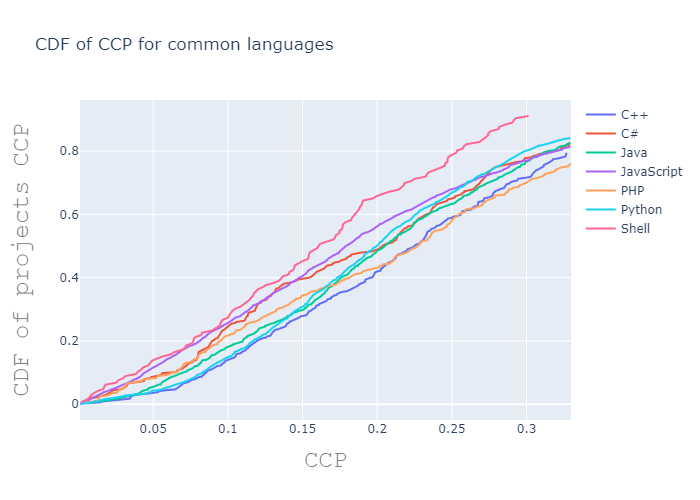}
\caption{\label{fig:ccp_per_lang} Cumulative distribution of CCP by language.  Distributions shifted to the right tend to have higher CCP.}
\end{figure}

We extracted the 100 most common file name extensions in GitHub, which cover 94\% of the files.
Of these, 28 extensions are of Turing-Complete programming languages (i.e., excluding languages like SQL).
We consider a language to be the dominant language in a project if above 80\% of files were in this language.
There were 5,407 projects with a dominant language out of the 7,557 being studied.
Figure \ref{fig:ccp_per_lang} shows the CDFs of the CCP of projects in major languages.

The figure focuses on the high to medium quality region (excluding the highest CCPs).
For averages see Table \ref{tab:speed_per_quality_and_ccp_per_lang}.
All languages cover a wide and overlapping range of CCP, and in all languages one can write high quality code.
The least bugs occurred in Shell.
This is an indication of the need to analyze quality carefully, as Shell is used to write scripts and should not be compared directly with languages used to write, for example, real-time applications.
Project in JavaScript, and to a somewhat lesser degree, in C\#, tend to have lower CCPs.
Higher CCPs occur in C++, and, towards the tail of the distribution, in PHP.
The rest of the languages are usually in between with changing regions of better performance. 

\hide{
\begin{table}\centering
\caption{\label{tab:speed_per_quality_and_ccp_per_lang}
CCP and development speed (commits per year of involved developers) per language. Values are averages $\pm$ standard errors.}
\begin{tabular}{| p{13mm}| c | p{12mm} | p{12mm}| p{12mm}|  p{12mm}|}
   \hline
   &   & \multicolumn{4}{c|}{Metric} \\ \cline{3-6}
Language & Projects & CCP & Speed & Speed in top 10\% & Speed in others \\ \hline
Shell & 146 & 17 $\pm$ 1.01 & 171 $\pm$ 10 & 185 $\pm$ 29 & 169 $\pm$ 11 \\ \hline
JavaScript & 1342 & 20 $\pm$ 0.38 & 156 $\pm$ 3 & 166 $\pm$ 8 & 154 $\pm$ 3 \\ \hline
C\# & 315 & 21 $\pm$ 0.78 & 181 $\pm$ 6 & 207 $\pm$ 27 & 178 $\pm$ 7 \\ \hline
Python & 1069 & 21 $\pm$ 0.39 & 139 $\pm$ 3 & 177 $\pm$ 19 & 137 $\pm$ 3 \\ \hline
Java & 764 & 22 $\pm$ 0.48 & 148 $\pm$ 4 & 205 $\pm$ 17 & 143 $\pm$ 4 \\ \hline
C++ & 341 & 23 $\pm$ 0.70 & 201 $\pm$ 7 & 324 $\pm$ 33 & 196 $\pm$ 7 \\ \hline
PHP & 326 & 24 $\pm$ 0.93 & 168 $\pm$ 6 & 180 $\pm$ 22 & 167 $\pm$ 6 \\ \hline
\end{tabular}
\end{table}
}
\begin{table}[h!]\centering
\caption{\label{tab:speed_per_quality_and_ccp_per_lang}
CCP and development speed (commits per year of involved developers) per language. Values are averages $\pm$ standard errors.}
\begin{tabular}{| p{13mm}| c | c | p{12mm}| p{12mm}|  p{12mm}|}
   \hline
   &   & \multicolumn{4}{c|}{Metric} \\ \cline{3-6}
Language & Projects & CCP & Speed & Speed in top 10\% & Speed in others \\ \hline
Shell & 146 & 0.18 $\pm$ 0.010 & 171 $\pm$ 10 & 185 $\pm$ 29 & 169 $\pm$ 11 \\ \hline
JavaScript & 1342 & 0.20 $\pm$ 0.004 & 156 $\pm$ 3 & 166 $\pm$ 8 & 154 $\pm$ 3 \\ \hline
C\# & 315 & 0.21 $\pm$ 0.008 & 181 $\pm$ 6 & 207 $\pm$ 27 & 178 $\pm$ 7 \\ \hline
Python & 1069 & 0.22 $\pm$ 0.004 & 139 $\pm$ 3 & 177 $\pm$ 19 & 137 $\pm$ 3 \\ \hline
Java & 764 & 0.22 $\pm$ 0.005 & 148 $\pm$ 4 & 205 $\pm$ 17 & 143 $\pm$ 4 \\ \hline
C++ & 341 & 0.24 $\pm$ 0.007 & 201 $\pm$ 7 & 324 $\pm$ 33 & 196 $\pm$ 7 \\ \hline
PHP & 326 & 0.25 $\pm$ 0.009 & 168 $\pm$ 6 & 180 $\pm$ 22 & 167 $\pm$ 6 \\ \hline
\end{tabular}
\end{table}

In order to verify that differences are not accidental, we split the projects by language and examined their average CCP.
An ANOVA test \cite{Fisher1919} led to a F-statistic of 8.3, indicating that language indeed has a substantial effect, with a p-value around $10^{-9}$.
Hence, as Table \ref{tab:speed_per_quality_and_ccp_per_lang} shows, there are statistically significant differences among the programming language, yet compared to the range of the CCP distribution they are small.

Of course, the above is not a full comparison of programming languages (See \cite{876288,7194625, Ray:2014:LSS:2635868.2635922, 10.1145/3340571} for comparisons and the difficulties involving them).
Many factors (e.g.\ being typed, memory allocation handing, compiled vs.\ dynamic) might cause the differences in the languages' CCP.
Our results agree with the results of \cite{Ray:2014:LSS:2635868.2635922, 10.1145/3340571}, indicating that difference between languages is usually small and that C++ has relatively high CCP.

\subsubsection{Developer Engagement and CCP}
\label{sect:CcpDeveloper Engagement}

The relation between churn (developers abandoning the project) and quality steps out of the technical field and involves human psychology.
Motivation influences performance \cite{campbell93,Wright00Satisfaction}.
Argyle investigated the relation between developers' happiness and their job satisfaction and work performance, showing ``modestly positive correlations with productivity, absenteeism and \emph{labour turnover}'' \cite{Argyle1989Happy}.
On the other direction, Ghayyur et al.\ conducted a survey in which 72\% claimed that poor code quality is demotivating \cite{Ghayyur2018}.
Hence, quality might be both the outcome and the cause of motivation.

\begin{figure}[h!]
\centering
\includegraphics[scale=0.35]{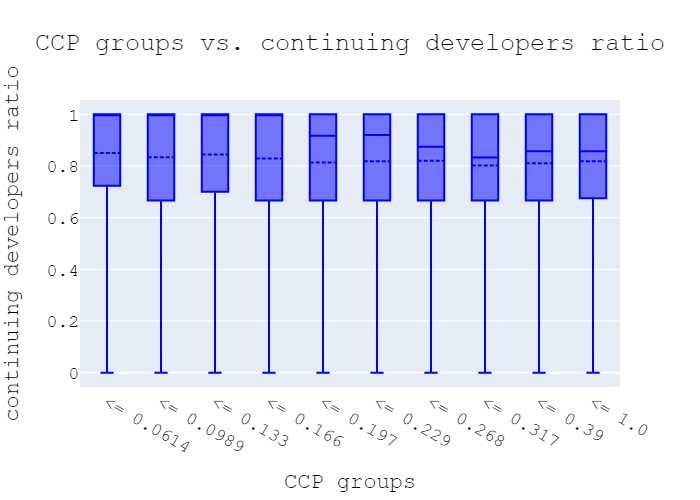}
\caption{\label{fig:continuing_developers_ratio}
Developer retention per CCP decile.
Note the change in the median.
}
\end{figure}

We checked the retention of involved developers, where retention is quantified as the percentage of developers that continue to work on the project in the next year, averaged over all years (Figure \ref{fig:continuing_developers_ratio}).
Note that the median is 100\% retention in all four top deciles, decreases over the next three, and stabilizes again at about 85\% in the last three CCP deciles.

When looking at co-change of CCP with churn ($1-retention$), the match is only 51\% for any change but 79\% for a change of at least 10 percentage points in each metric.
A improvement of 10 percent points in CCP leads to a significant improvement in churn in 21\% of the cases, a lift of 17\%.
When controlling the language, age group, or developer number group, we still get matching co-change.

\begin{figure}[h!]
\centering
\includegraphics[scale=0.35]{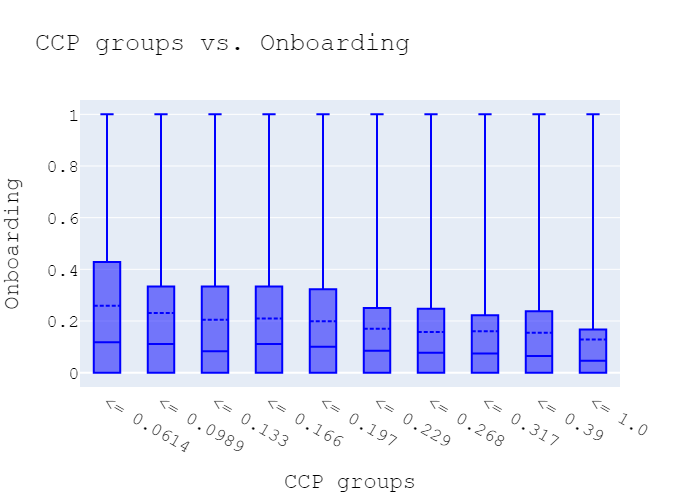}
\caption{\label{fig:Onboarding}
On-boarding per CCP decile.
}
\end{figure}

acquiring new developers complements the retention of existing ones.
We define the on-boarding ratio as the average percentage of new developers becoming involved.
Figure \ref{fig:Onboarding} shows that the higher the CCP, the lower is the on-boarding, and on-boarding average is doubled in the first decile compared to the last.
In order to be more robust to noise, we consider projects that have at least 10 new developers.
When looking at co-change of on-boarding and CCP, the match is only 53\% for any change but 85\% for a change of at least 10 percent points in both metrics.
An improvement of 10 percent points in CCP leads to a significant improvement in on-boarding in 10\% of the cases, a lift of 18\%.
When controlling on language, results fit the relation other than in PHP and Shell (which had a small number of cases).
Results hold for all age groups.
For size, they hold for intermediate and numerous numbers of developers; by definition, with few developers there are no projects with at least 10 new developers.

\subsubsection{Development Speed and CCP}
\label{sect:CcpSpeed}

Like quality, the definition of productivity is subjective and not clear.
Measures including LOC \cite{544349}, modules \cite{960633}, and function points \cite{820015,Jiang2007Productivity} per time unit have been suggested and criticized \cite{Kemerer:1992:IRF:141344.141614, Kemerer:1993:RFP:151220.151230}.
We chose to measure development speed by the number of commits per developer per year.
This is an output per time measure, and the inverse of time to complete a task, investigated in the classical work of Sackman et al. \cite{Sackman:1968:EES:362851.362858}.
The number of commits is correlated with self-rated productivity\cite{47853} and team lead perception of productivity \cite{Oliveira2020TLProd}.
We chose commits as the output unit since a commit is a unit of work, its computation is easy and objective, and it is not biased toward implementation details.

The number of commits per project per year is stable with a Pearson correlation of 0.71.
The number of developers per year is also stable with a Pearson correlation 0.81.
To study development speed we omit developers with fewer than 12 commits per year since they are non-involved developers.
We also capped the number of commits per developer at 500, about the 99th percentile of the developers' contributions.
While commits by users below the 99th percentile are only 73\% of the total, excluding the long tail (which reaches 300,000 commits) is justified because it most probably does not represent usual manual human effort.
Using both restrictions the correlation of commits per developer in adjacent years is 0.62 (compared to 0.59 without them), which is reasonably stable.

\begin{figure}[h!]
\centering
\includegraphics[scale=0.35]{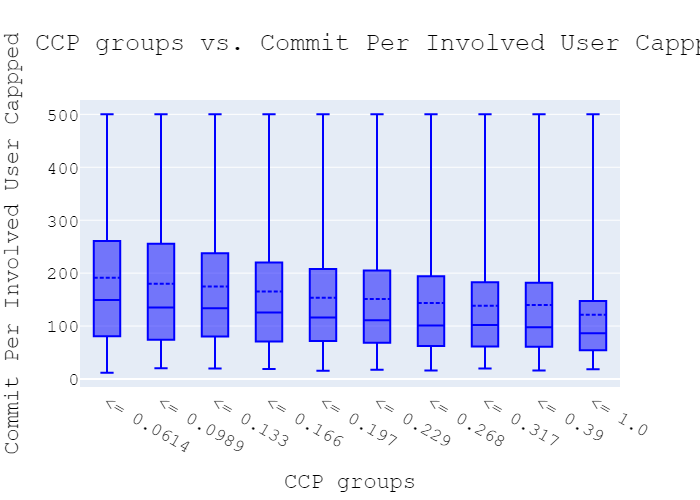}
\caption{\label{fig:commit_per_user_above_11_cap}
Distribution of commits of involved developers (capped) per CCP decile.
}
\end{figure}

As Figure \ref{fig:commit_per_user_above_11_cap} shows, there is a steady decrease of speed with CCP.
The average speed in the first decile is 56\% higher than in the last one.
As with file length, speed differs in projects written in different languages.
Yet in all of them higher quality goes with higher speed (see Table \ref{tab:speed_per_quality_and_ccp_per_lang}).

\hide{
\begin{figure}[h!]
\centering
\includegraphics[scale=0.4,trim={0mm 20mm 0 30mm},clip]{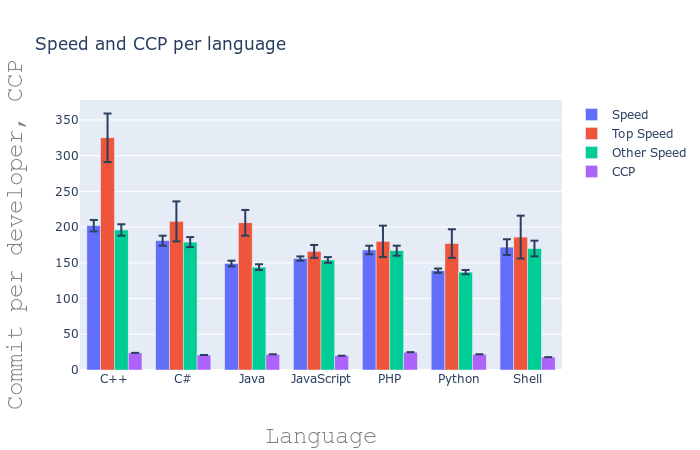}
\caption{\label{fig:speed_per_quality_and_ccp_per_lang}
CCP and Speed per language and quality group. Error bar is 1 std. error.}
\end{figure}
}

We conducted twin experiments, to control the developer.
When a developer works in a faster project, he is faster than himself in other projects in 51\% of the cases, 8\% lift.
When the project speed is 10 commits larger, the developer has 42\% chance to be also 10 commits faster than himself, a lift of 11\%.

We also investigated the co-change of CCP and speed.
In 52\% of the cases, an improvement in CCP goes with improvement in speed.
Given a CCP improvement, there is a speed improvement in 53\% of the cases, a lift of 4\%.
Given a 10 percent points improvement in CCP, the probability of 10 more commits per year per developer is 53\%, and the lift is 2\%.
In the other direction, given an improvement in speed the probability of a significant improvement in CCP drops to 7\%.
Hence, knowing of a significant improvement in CCP a speed improvement is likely, but knowing of a speed improvement a significant CCP improvement is very unlikely.

When controlling for age or language, results hold.
Results also hold for intermediate and numerous developer groups, with a positive lift when the change is significant, but a -3\% lift in the few developers group for any change.

There are two differing theories regarding the relations between quality and productivity.
The classical Iron Triangle \cite{IronTriangle} sees them as a trade-off: investment in quality comes at the expense of productivity.
On the other hand, ``Quality is Free'' claims that investment in quality is beneficial and leads to increased productivity \cite{crosby:1979:free_quality}.
Our results in Table \ref{tab:speed_per_quality_and_ccp_per_lang} enable a quantitative investigation of this issue, where speed is operationalized by commits per year and quality by CCP.

As ``Quality is Free'' predicts, we find that in high quality projects the development speed is much higher.
The twin experiments help to reduce noise, demonstrating that development speed is a characteristic of the project.
In case that this correlation is indeed due to causality, then when you improve the quality you also gain speed, enjoying both worlds.
This relation between quality and development speed is also supported by Jones's research on time wasted due to low quality \cite{Jones2015Wastage, Jones2006Reasons} and developers performing the same tasks during ``Personal Software Process'' training \cite{nc2020assessing}.

\section{Threats to Validity}
\label{sect:CcpThreats}

There is no agreed quantitative definition of quality, hence we cannot ensure that a certain metric measures quality.
In order to cope with this, we showed that our metric agrees with developers' comments on quality and is associated with variables that are believed to reflect or influence quality.

A specific threat to validity in our work is related to construct validity.
We set out to measure the Corrective Commit Probability and do so based on a linguistic analysis.
We investigated whether it is indeed \hyperref[sect:MleValidation]{accurate and precise} in Section \ref{sect:MleValidation}.

The number of test labeled commits is small, about 1,000, hence there is a question of how well they represent the underlying distribution.
We evaluated the sensitivity to changes in the data. 
Since the model was built mainly using domain knowledge and a different data set, we could use a small training set.
Therefore, we preferred to use most of the labels as test set for the variables estimation and to improve the estimation of the $recall$ and $Fpr$.

The labeling was done manually by humans who are prone to error and subjectivity.
In order to make the labeling stricter, we used a labeling protocol.
Out of the samples, 400 were labeled by three annotators independently.
The labels were compared in order to evaluate the amount of uncertainty. 

Other than uncertainty due to different opinions, there was uncertainty due to the lack of information in the commit message. 
For example, the message ``Changed result default value to False'' describes a change well but leaves us uncertain regarding its nature.
We used the \hyperref[sect:GoldStandard]{gold standard} labels to verify that this is rare.

Our main assumption is the conditional independence between the corrective commits (code) and the commit messages describing them (process) given our concept (the commit being corrective, namely a bug fix). 
This means that the model performance is the same over all the projects, and a different hit rate is due to a different CCP.
This assumption is invalid in some cases. 
For example, projects documented in a language other than English will appear to have no bugs.
Non-English commit messages are relatively easy to identify; more problematic are differences in English fluency.
Native English speakers are less likely to have spelling mistakes and typos.
A spelling mistake might prevent our model from identifying the textual pattern, thus lowering the recall.
This will lead to an illusive benefit of spelling mistakes, misleading us to think that people who tend to have more spelling mistakes tend to have fewer bugs.

Another threat to validity is due to the family of models that we chose to use.
We chose to represent the model using two parameters, $recall$ and $Fpr$, following the guidance of Occam's razor and resorting to a more complex solution only when a need arises.
However, many other families of models are possible.
We could consider different sub-models for various message lengths, a model that predicts the commit category instead of the Boolean "Is Corrective" concept, etc.\
Each family will have different parameters and behavior.
More complex models will have more representative power but will be harder to learn and require more samples.

A common assumption in statistical analysis is the IID assumption (Independent and Identically Distributed random variables).
This assumption clearly doesn't hold for GitHub projects. 
We found that forks, projects based on others and sharing a common history, were 35\% of the active projects.
We therefore removed forks, but projects might still share code and commits.
Also, older projects, with more commits and users, have higher weight in twin studies and co-change analysis.

Our metric focuses on the fraction of commits that \emph{correct} bugs.
One can claim that the fraction of commits that \emph{induce} bugs is a better quality-metric.
In principle, this can be done using the SZZ algorithm (the common algorithm for identifying bug inducing commits \cite{Sliwerski:2005:CIF:1082983.1083147}).
But note that SZZ is applied after the bug was identified and fixed.
Thus, the inducing and fixing commits are actually expected to give similar results.

Another major threat concerns internal validity.
Our basic assumption is that corrective commits reflect bugs, and therefore a low CCP is indicative of few bugs and high quality.
But a low CCP can also result from a disregard for fixing bugs or an inability to do so.
On the other hand, in extremely popular projects, Linus's law ``given enough eyeballs, all bugs are shallow'' \cite{LinusRule} might lead to more effective bug identification and high CCP.
Another unwanted effect of using corrective commits is that improvements in bug detection (e.g., by doubling the QA department) will look like a reduction in quality.
The correlations found between CCP and various other indicators of software quality add confidence that CCP is indeed valid.
We identify such cases and discuss them in  \hyperref[sect:DetectionEfficiency]{Section \ref{sect:DetectionEfficiency}}.

Focusing on corrective commits also leads to several biases.
Most obviously, existing bugs that have not been found yet are unknown.
Finding and fixing bugs might take months \cite{Kim:2006:LDT:1137983.1138027}.
When projects differ in the time needed to identify a bug, our results will be biased.

Software development is usually done subject to lack of time and resources. 
Due to that, many times known bugs of low severity are not fixed.
While this leads to a bias, it can be considered to be a desirable one, by focusing on the more important bugs.

A threat to external validity might arise due to the use of open source projects that might not represent projects done in software companies.
We feel that the open source projects are of significant interest on their own.
Other than that, the projects we analyzed include projects of Google, Microsoft, Apple, etc.\ so at least part of the area is covered.

Time, cost, and development speed are problematic to measure.
We use commits as a proxy to work since they typically represent tasks.
Yet, tasks differ in size and difficulty and their translation to commits might differ due to the project or developer habits.
Commits may also include a mix of different tasks.
In order to reduce the influence of project culture we aggregated many of them.
In order to eliminate the effect of personal habits, we used twins experiments.
Other than that, the number of commits per time is correlated to developers' self-rated productivity \cite{47853} and team lead perception of productivity \cite{Oliveira2020TLProd}, hence it provides a good computable estimator.

\section{Conclusions}

We presented a novel way to measure projects' code quality, using the Corrective Commit Probability (CCP).
We use the consensus that bugs are bad and indicative of low quality to base a metric on them.
We started off with a linguistic model to identify corrective commits, significantly improving prior work \cite{5090025,Amor_discriminatingdevelopment,Levin:2017:BAC:3127005.3127016, Amit:2019:RRB:3345629.3345631},
and developed a mathematical method to find the most likely CCP given the model's hit rate.
The CCP metric has the following properties:
\begin{itemize}
\item It matches developers' references to quality.
\item It is stable: it reflects the character of a project and does not change much from year to year.
\item It is informative in that it has a wide range of values and distinguishes between projects.
\end{itemize}

We estimated the CCP of all 7,557 large active projects in BigQuery's GitHub data.
This created a quality scale, enabling observations on the state of the practice.
Using this scale developers can compare their project's quality (as reflected by CCP) to the community.
A low percentile suggests the need for quality improvement efforts.

We checked the sensitivity of our assumptions and noticed that the theoretical invalid CCP range indeed tend to be not in English or not software.
A difference in bug detection efficiency was demonstrated in highly popular projects, supporting Linus's law \cite{LinusRule}.

Our results also helped demonstrate that ``Quality is Free''.
Instead of a trade-off between quality and development speed, we find that they are positively correlated, and this was further supported with co-change analysis and twin experiments.
Thus, investing in quality may actually reduce schedules rather than extending them. 

We show a correlation between short files and low coupling and quality, supporting to well-know recommendation for quality improvement.
Hence, if the discussed relations are indeed casual, we have a simple way to reach high quality, which will benefit a project also in higher productivity, better on-boarding, and lower churn.

\subsection*{Supplementary Materials}

The language models are available at \url{https://github.com/evidencebp/commit-classification}
Utilities used for the analysis (e.g., co-change) are at \url{https://github.com/evidencebp/analysis_utils}
All other supplementary materials can be found at \url{https://github.com/evidencebp/corrective-commit-probability}

\subsection*{Acknowledgements}

This research was supported by \anon{the ISRAEL SCIENCE FOUNDATION (grant No.\ 832/18)}{[anonymized]}.
We thank Amiram Yehudai and Stanislav Levin for providing us their data set of labeled commits \cite{Levin:2017:BAC:3127005.3127016}.
We thank Guilherme Avelino for drawing our attention to the importance of Truck Factor Developers Detachment (TFDD) and providing a data set \cite{DBLP:journals/corr/abs-1906-08058}.

\bibliographystyle{abbrv}
\bibliography{abbrv.bib,se.bib,bibtex.bib}

\end{document}

%% file: commands.tex
\usepackage{enumitem}
\usepackage{subfigure}
\usepackage{amsmath}
\usepackage{euscript}
\usepackage{centernot}
\usepackage{multirow}
\usepackage[]{algorithm2e}

\usepackage{array}
\newcolumntype{L}[1]{>{\raggedright\let\newline\\\arraybackslash\hspace{0pt}}m{#1}}
\newcolumntype{C}[1]{>{\centering\let\newline\\\arraybackslash\hspace{0pt}}m{#1}}
\newcolumntype{R}[1]{>{\raggedleft\let\newline\\\arraybackslash\hspace{0pt}}m{#1}}

\usepackage{tikz}
\usetikzlibrary{arrows}
\usetikzlibrary{shapes}

\newcommand{\hide}[1]{}
\newcommand{\comment}[1]{}

\newif\ifshowcomment
\showcommenttrue
\ifshowcomment
    \newcommand{\anon}[2]{#1}

    \newcommand{\dror}[1]{\textsf{\color{purple}{[{Dror: #1}]}}}
    \newcommand{\idan}[1]{\textsf{\color{violet}{[{Idan: #1}]}}}

\else
    \newcommand{\anon}[2]{#2}
    
    \newcommand{\dror}[1]{}
    \newcommand{\idan}[1]{}
\fi

\newcommand{\remove}[1]{}

%% file: main.bbl
\begin{thebibliography}{100}

\bibitem{alkilidar05}
H.~Al-Kilidar, K.~Cox, and B.~Kitchenham.
\newblock The use and usefulness of the {ISO/IEC} 9126 quality standard.
\newblock In {\em Intl.\ Synp.\ Empirical Softw.\ Eng.}, pages 126--132, Nov
  2005.

\bibitem{10.1145/3359591.3359735}
M.~Allamanis.
\newblock The adverse effects of code duplication in machine learning models of
  code.
\newblock In {\em Proceedings of the 2019 ACM SIGPLAN International Symposium
  on New Ideas, New Paradigms, and Reflections on Programming and Software},
  Onward! 2019, page 143–153, New York, NY, USA, 2019. Association for
  Computing Machinery.

\bibitem{Amit:2019:RRB:3345629.3345631}
I.~Amit and D.~G. Feitelson.
\newblock Which refactoring reduces bug rate?
\newblock In {\em Proceedings of the Fifteenth International Conference on
  Predictive Models and Data Analytics in Software Engineering}, PROMISE'19,
  pages 12--15, New York, NY, USA, 2019. ACM.

\bibitem{archimedes}
I.~Amit, E.~Firstenberg, and Y.~Meshi.
\newblock Framework for semi-supervised learning when no labeled data is given.
\newblock U.S. patent application \#US20190164086A1, 2017.

\bibitem{Amor_discriminatingdevelopment}
J.~J. Amor, G.~Robles, J.~M. Gonzalez-Barahona, and A.~Navarro.
\newblock Discriminating development activities in versioning systems: A case
  study, Jan 2006.

\bibitem{10.1145/1463788.1463819}
G.~Antoniol, K.~Ayari, M.~Di~Penta, F.~Khomh, and Y.-G. Gu\'{e}h\'{e}neuc.
\newblock Is it a bug or an enhancement? a text-based approach to classify
  change requests.
\newblock In {\em Proceedings of the 2008 Conference of the Center for Advanced
  Studies on Collaborative Research: Meeting of Minds}, CASCON ’08, New York,
  NY, USA, 2008. Association for Computing Machinery.

\bibitem{Argyle1989Happy}
M.~Argyle.
\newblock Do happy workers work harder? the effect of job satisfaction on job
  performance.
\newblock In R.~Veenhoven, editor, {\em How harmful is happiness? Consequences
  of enjoying life or not}. Universitaire Pers, Rotterdam, The Netherlands,
  1989.

\bibitem{DBLP:journals/corr/abs-1906-08058}
G.~Avelino, E.~Constantinou, M.~T. Valente, and A.~Serebrenik.
\newblock On the abandonment and survival of open source projects: An empirical
  investigation.
\newblock {\em CoRR}, abs/1906.08058, 2019.

\bibitem{DBLP:journals/corr/AvelinoPHV16}
G.~Avelino, L.~T. Passos, A.~C. Hora, and M.~T. Valente.
\newblock A novel approach for estimating truck factors.
\newblock {\em CoRR}, abs/1604.06766, 2016.

\bibitem{544352}
V.~R. Basili, L.~C. Briand, and W.~L. Melo.
\newblock A validation of object-oriented design metrics as quality indicators.
\newblock {\em IEEE Transactions on Software Engineering}, 22(10):751--761, Oct
  1996.

\bibitem{bavota15}
G.~Bavota, A.~De~Lucia, M.~Di~Penta, R.~Oliveto, and F.~Palomba.
\newblock An experimental investigation on the innate relationship between
  quality and refactoring.
\newblock {\em J.\ Syst.\ \& Softw.}, 107:1--14, Sep 2015.

\bibitem{10.1145/3340571}
E.~D. Berger, C.~Hollenbeck, P.~Maj, O.~Vitek, and J.~Vitek.
\newblock On the impact of programming languages on code quality: A
  reproduction study.
\newblock {\em ACM Trans. Program. Lang. Syst.}, 41(4), Oct 2019.

\bibitem{Bird:2009:FBB:1595696.1595716}
C.~Bird, A.~Bachmann, E.~Aune, J.~Duffy, A.~Bernstein, V.~Filkov, and
  P.~Devanbu.
\newblock Fair and balanced?: Bias in bug-fix datasets.
\newblock In {\em Proceedings of the the 7th Joint Meeting of the European
  Software Engineering Conference and the ACM SIGSOFT Symposium on The
  Foundations of Software Engineering}, ESEC/FSE '09, pages 121--130, New York,
  NY, USA, 2009. ACM.

\bibitem{bird2011don}
C.~Bird, N.~Nagappan, B.~Murphy, H.~Gall, and P.~Devanbu.
\newblock Don't touch my code! examining the effects of ownership on software
  quality.
\newblock In {\em Proceedings of the 19th ACM SIGSOFT symposium and the 13th
  European conference on Foundations of software engineering}, pages 4--14,
  2011.

\bibitem{5069475}
C.~Bird, P.~C. Rigby, E.~T. Barr, D.~J. Hamilton, D.~M. German, and P.~Devanbu.
\newblock The promises and perils of mining git.
\newblock In {\em 2009 6th IEEE International Working Conference on Mining
  Software Repositories}, pages 1--10, May 2009.

\bibitem{Boehm:2001:SDR:619059.621640}
B.~Boehm and V.~R. Basili.
\newblock Software defect reduction top 10 list.
\newblock {\em Computer}, 34(1):135--137, Jan 2001.

\bibitem{boehm:econo}
B.~W. Boehm.
\newblock {\em Software Engineering Economics}.
\newblock Prentice-Hall, 1981.

\bibitem{boehm76}
B.~W. Boehm, J.~R. Brown, and M.~Lipow.
\newblock Quantitative evaluation of software quality.
\newblock In {\em Intl.\ Conf.\ Softw.\ Eng.}, number~2, pages 592--605, Oct
  1976.

\bibitem{6191}
B.~W. Boehm and P.~N. Papaccio.
\newblock Understanding and controlling software costs.
\newblock {\em IEEE Transactions on Software Engineering}, 14(10):1462--1477,
  Oct 1988.

\bibitem{BOX1979201}
G.~Box.
\newblock Robustness in the strategy of scientific model building.
\newblock In R.~L. LAUNER and G.~N. WILKINSON, editors, {\em Robustness in
  Statistics}, pages 201 -- 236. Academic Press, 1979.

\bibitem{brooks:book}
F.~P. Brooks, Jr.
\newblock {\em The Mythical Man-Month: Essays on Software Engineering}.
\newblock Addison-Wesley, 1975.

\bibitem{campbell93}
J.~P. Campbell, R.~A. McCloy, S.~H. Oppler, and C.~E. Sager.
\newblock A theory of performance.
\newblock In N.~Schmitt, W.~C. Borman, and Associates, editors, {\em Personnel
  Selection in Organizations}, pages 35--70. Jossey-Bass Pub., 1993.

\bibitem{Chidamber:1994:MSO:630808.631131}
S.~R. Chidamber and C.~F. Kemerer.
\newblock A metrics suite for object oriented design.
\newblock {\em IEEE Trans. Softw. Eng.}, 20(6):476--493, Jun 1994.

\bibitem{Cohens1960Kappa}
J.~Cohen.
\newblock A coefficient of agreement for nominal scales.
\newblock {\em Educational and Psychological Measurement}, 20(1):37--46, 1960.

\bibitem{crosby:1979:free_quality}
P.~Crosby.
\newblock {\em Quality Is Free: The Art of Making Quality Certain}.
\newblock McGrawHill, 1979.

\bibitem{10.1145/157709.157715}
W.~Cunningham.
\newblock The wycash portfolio management system.
\newblock In {\em Addendum to the Proceedings on Object-Oriented Programming
  Systems, Languages, and Applications (Addendum)}, OOPSLA ’92, page 29–30,
  New York, NY, USA, 1992. Association for Computing Machinery.

\bibitem{5463279}
M.~D'Ambros, M.~Lanza, and R.~Robbes.
\newblock An extensive comparison of bug prediction approaches.
\newblock In {\em 2010 7th IEEE Working Conference on Mining Software
  Repositories (MSR 2010)}, pages 31--41, May 2010.

\bibitem{10.2307/2346806}
A.~P. Dawid and A.~M. Skene.
\newblock Maximum likelihood estimation of observer error-rates using the em
  algorithm.
\newblock {\em Journal of the Royal Statistical Society. Series C (Applied
  Statistics)}, 28(1):20--28, 1979.

\bibitem{Dawson2010SDLC}
M.~Dawson, D.~Burrell, E.~Rahim, and S.~Brewster.
\newblock Integrating software assurance into the software development life
  cycle ({SDLC}).
\newblock {\em Journal of Information Systems Technology and Planning},
  3:49--53, 2010.

\bibitem{dromey95}
G.~Dromey.
\newblock A model for software product quality.
\newblock {\em IEEE Trans.\ Softw.\ Eng.}, 21(2):146--162, Feb 1995.

\bibitem{Efron1992}
B.~Efron.
\newblock {\em Bootstrap Methods: Another Look at the Jackknife}, pages
  569--593.
\newblock Springer New York, New York, NY, 1992.

\bibitem{Fisher1919}
R.~Fisher.
\newblock The correlation between relatives on the supposition of mendelian
  inheritance.
\newblock {\em Transactions of the Royal Society of Edinburgh}, 52(2):399--433,
  1919.

\bibitem{1999:RID:311424}
M.~Fowler.
\newblock {\em Refactoring: Improving the Design of Existing Code}.
\newblock Addison-Wesley Longman Publishing Co., Inc., Boston, MA, USA, 1999.

\bibitem{fowler1997refactoring}
M.~Fowler, K.~Beck, and W.~R. Opdyke.
\newblock Refactoring: Improving the design of existing code.
\newblock In {\em 11th European Conference. Jyv{\"a}skyl{\"a}, Finland}, 1997.

\bibitem{Gharehyazie2019}
M.~Gharehyazie, B.~Ray, M.~Keshani, M.~S. Zavosht, A.~Heydarnoori, and
  V.~Filkov.
\newblock Cross-project code clones in github.
\newblock {\em Empirical Software Engineering}, 24(3):1538--1573, Jun 2019.

\bibitem{Ghayyur2018}
S.~A.~K. Ghayyur, S.~Ahmed, S.~Ullah, and W.~Ahmed.
\newblock The impact of motivator and demotivator factors on agile software
  development.
\newblock {\em International Journal of Advanced Computer Science and
  Applications}, 9(7), 2018.

\bibitem{gil17}
Y.~Gil and G.~Lalouche.
\newblock On the correlation between size and metric validity.
\newblock {\em Empirical Softw.\ Eng.}, 22(5):2585--2611, Oct 2017.

\bibitem{859533}
T.~L. Graves, A.~F. Karr, J.~S. Marron, and H.~Siy.
\newblock Predicting fault incidence using software change history.
\newblock {\em IEEE Transactions on Software Engineering}, 26(7):653--661, July
  2000.

\bibitem{1542070}
T.~Gyimothy, R.~Ferenc, and I.~Siket.
\newblock Empirical validation of object-oriented metrics on open source
  software for fault prediction.
\newblock {\em IEEE Transactions on Software Engineering}, 31(10):897--910, Oct
  2005.

\bibitem{hackbarth16}
R.~Hackbarth, A.~Mockus, J.~Palframan, and R.~Sethi.
\newblock Improving software quality as customers perceive it.
\newblock {\em IEEE Softw.}, 33(4):40--45, Jul/Aug 2016.

\bibitem{Hall:2012:SLR:2420627.2420790}
T.~Hall, S.~Beecham, D.~Bowes, D.~Gray, and S.~Counsell.
\newblock A systematic literature review on fault prediction performance in
  software engineering.
\newblock {\em IEEE Trans. Softw. Eng.}, 38(6):1276--1304, Nov 2012.

\bibitem{Halstead:1977:ESS:540137}
M.~H. Halstead.
\newblock {\em Elements of Software Science (Operating and Programming Systems
  Series)}.
\newblock Elsevier Science Inc., New York, NY, USA, 1977.

\bibitem{hastings1947}
C.~Hastings, F.~Mosteller, J.~W. Tukey, and C.~P. Winsor.
\newblock Low moments for small samples: A comparative study of order
  statistics.
\newblock {\em Ann. Math. Statist.}, 18(3):413--426, 09 1947.

\bibitem{Herzig:2013:IBI:2486788.2486840}
K.~Herzig, S.~Just, and A.~Zeller.
\newblock It's not a bug, it's a feature: How misclassification impacts bug
  prediction.
\newblock In {\em Proceedings of the 2013 International Conference on Software
  Engineering}, ICSE '13, pages 392--401, Piscataway, NJ, USA, 2013. IEEE
  Press.

\bibitem{5090025}
A.~{Hindle}, D.~M. {German}, M.~W. {Godfrey}, and R.~C. {Holt}.
\newblock Automatic classication of large changes into maintenance categories.
\newblock In {\em 2009 IEEE 17th International Conference on Program
  Comprehension}, pages 30--39, May 2009.

\bibitem{Hovemeyer:2004:FBE:1052883.1052895}
D.~Hovemeyer and W.~Pugh.
\newblock Finding bugs is easy.
\newblock {\em SIGPLAN Not.}, 39(12):92--106, Dec 2004.

\bibitem{iec20019126}
I.~IEC.
\newblock 9126-1 (2001). software engineering product quality-part 1: Quality
  model.
\newblock {\em International Organization for Standardization}, page~16, 2001.

\bibitem{ISO}
{International Organization for Standardization}.
\newblock Systems and software engineering -- systems and software quality
  requirements and evaluation (square) -- system and software quality models,
  2011.

\bibitem{Jiang2007Productivity}
Z.~Jiang, P.~Naudé, and C.~Comstock.
\newblock An investigation on the variation of software development
  productivity.
\newblock {\em IEEE Transactions on Software Engineering}, 1(2):72--81, 2007.

\bibitem{Jones:1991:ASM:109758}
C.~Jones.
\newblock {\em Applied Software Measurement: Assuring Productivity and
  Quality}.
\newblock McGraw-Hill, Inc., New York, NY, USA, 1991.

\bibitem{Jones2006Reasons}
C.~Jones.
\newblock Social and technical reasons for software project failures.
\newblock {\em CrossTalk, The J. Def. Software Eng.}, 19(6):4--9, 2006.

\bibitem{CapersJonesQuality2012}
C.~Jones.
\newblock Software quality in 2012: A survey of the state of the art, 2012.
\newblock [Online; accessed 24-September-2018].

\bibitem{Jones2015Wastage}
C.~Jones.
\newblock Wastage: The impact of poor quality on software economics. retrieved
  from http://asq.org/pub/sqp/.
\newblock {\em Software Quality Professional}, 18(1):23–32, 2015.

\bibitem{Kalliamvakou15Promises}
E.~Kalliamvakou, G.~Gousios, K.~Blincoe, L.~Singer, D.~German, and D.~Damian.
\newblock The promises and perils of mining github (extended version).
\newblock {\em Empirical Software Engineering}, 01 2015.

\bibitem{6341763}
Y.~Kamei, E.~Shihab, B.~Adams, A.~E. Hassan, A.~Mockus, A.~Sinha, and
  N.~Ubayashi.
\newblock A large-scale empirical study of just-in-time quality assurance.
\newblock {\em IEEE Transactions on Software Engineering}, 39(6):757--773, June
  2013.

\bibitem{Kemerer:1993:RFP:151220.151230}
C.~F. Kemerer.
\newblock Reliability of function points measurement: A field experiment.
\newblock {\em Commun. ACM}, 36(2):85--97, Feb 1993.

\bibitem{Kemerer:1992:IRF:141344.141614}
C.~F. Kemerer and B.~S. Porter.
\newblock Improving the reliability of function point measurement: An empirical
  study.
\newblock {\em IEEE Trans. Softw. Eng.}, 18(11):1011--1024, Nov 1992.

\bibitem{Khomh:2012:FRI:2664446.2664475}
F.~Khomh, T.~Dhaliwal, Y.~Zou, and B.~Adams.
\newblock Do faster releases improve software quality?: An empirical case study
  of mozilla firefox.
\newblock In {\em Proceedings of the 9th IEEE Working Conference on Mining
  Software Repositories}, MSR '12, pages 179--188, Piscataway, NJ, USA, 2012.
  IEEE Press.

\bibitem{khomh2009exploratory}
F.~Khomh, M.~Di~Penta, and Y.-G. Gueheneuc.
\newblock An exploratory study of the impact of code smells on software
  change-proneness.
\newblock In {\em 2009 16th Working Conference on Reverse Engineering}, pages
  75--84. IEEE, 2009.

\bibitem{Kim:2006:LDT:1137983.1138027}
S.~Kim and E.~J. Whitehead, Jr.
\newblock How long did it take to fix bugs?
\newblock In {\em Proceedings of the 2006 International Workshop on Mining
  Software Repositories}, MSR '06, pages 173--174, New York, NY, USA, 2006.
  ACM.

\bibitem{Kim:2007:PFC:1248820.1248881}
S.~Kim, T.~Zimmermann, E.~J. Whitehead~Jr., and A.~Zeller.
\newblock Predicting faults from cached history.
\newblock In {\em Proceedings of the 29th International Conference on Software
  Engineering}, ICSE '07, pages 489--498, Washington, DC, USA, 2007. IEEE
  Computer Society.

\bibitem{kruchten2012technical}
P.~Kruchten, R.~L. Nord, and I.~Ozkaya.
\newblock Technical debt: From metaphor to theory and practice.
\newblock {\em Ieee software}, 29(6):18--21, 2012.

\bibitem{1456074}
M.~M. {Lehman}.
\newblock Programs, life cycles, and laws of software evolution.
\newblock {\em Proceedings of the IEEE}, 68(9):1060--1076, 1980.

\bibitem{lehman97}
M.~M. Lehman, J.~F. Ramil, P.~D. Wernick, D.~E. Perry, and W.~M. Turski.
\newblock Metrics and laws of software evolution -- the nineties view.
\newblock In {\em Intl.\ Software Metrics Symp.}, number~4, pages 20--32, Nov
  1997.

\bibitem{Levin:2017:BAC:3127005.3127016}
S.~Levin and A.~Yehudai.
\newblock Boosting automatic commit classification into maintenance activities
  by utilizing source code changes.
\newblock In {\em Proceedings of the 13th International Conference on
  Predictive Models and Data Analytics in Software Engineering}, PROMISE, pages
  97--106, New York, NY, USA, 2017. ACM.

\bibitem{lientz78}
B.~P. Lientz, E.~B. Swanson, and G.~E. Tompkins.
\newblock Characteristics of application software maintenance.
\newblock {\em Comm.\ ACM}, 21(6):466--471, Jun 1978.

\bibitem{lipow1982number}
M.~Lipow.
\newblock Number of faults per line of code.
\newblock {\em IEEE Transactions on software Engineering}, (4):437--439, 1982.

\bibitem{10.1145/3133908}
C.~V. Lopes, P.~Maj, P.~Martins, V.~Saini, D.~Yang, J.~Zitny, H.~Sajnani, and
  J.~Vitek.
\newblock D\'{e}j\`{a}vu: A map of code duplicates on github.
\newblock {\em Proc. ACM Program. Lang.}, 1(OOPSLA), Oct 2017.

\bibitem{820015}
K.~D. {Maxwell} and P.~{Forselius}.
\newblock Benchmarking software development productivity.
\newblock {\em IEEE Software}, 17(1):80--88, Jan 2000.

\bibitem{544349}
K.~D. {Maxwell}, L.~{Van Wassenhove}, and S.~{Dutta}.
\newblock Software development productivity of european space, military, and
  industrial applications.
\newblock {\em IEEE Transactions on Software Engineering}, 22(10):706--718, Oct
  1996.

\bibitem{McCabe:1976:CM:1313324.1313586}
T.~J. McCabe.
\newblock A complexity measure.
\newblock {\em IEEE Trans. Softw. Eng.}, 2(4):308--320, Jul 1976.

\bibitem{mockus2020complete}
A.~Mockus, D.~Spinellis, Z.~Kotti, and G.~J. Dusing.
\newblock A complete set of related git repositories identified via community
  detection approaches based on shared commits, 2020.

\bibitem{10.1007/978-3-030-40223-5_8}
A.-J. Molnar, A.~Neam{\c{T}}u, and S.~Motogna.
\newblock Evaluation of software product quality metrics.
\newblock In E.~Damiani, G.~Spanoudakis, and L.~A. Maciaszek, editors, {\em
  Evaluation of Novel Approaches to Software Engineering}, pages 163--187,
  Cham, 2020. Springer International Publishing.

\bibitem{960633}
S.~{Morasca} and G.~{Russo}.
\newblock An empirical study of software productivity.
\newblock In {\em 25th Annual International Computer Software and Applications
  Conference. COMPSAC 2001}, pages 317--322, Oct 2001.

\bibitem{Moser:2008:ARS:1414004.1414063}
R.~Moser, W.~Pedrycz, and G.~Succi.
\newblock Analysis of the reliability of a subset of change metrics for defect
  prediction.
\newblock In {\em Proceedings of the Second ACM-IEEE International Symposium on
  Empirical Software Engineering and Measurement}, ESEM '08, pages 309--311,
  New York, NY, USA, 2008. ACM.

\bibitem{Munaiah17Curating}
N.~Munaiah, S.~Kroh, C.~Cabrey, and M.~Nagappan.
\newblock Curating github for engineered software projects.
\newblock {\em Empirical Software Engineering}, 22, 04 2017.

\bibitem{47853}
E.~Murphy-Hill, C.~Jaspan, C.~Sadowski, D.~C. Shepherd, M.~Phillips, C.~Winter,
  A.~K. Dolan, E.~K. Smith, and M.~A. Jorde.
\newblock What predicts software developers’ productivity?
\newblock {\em Transactions on Software Engineering}, 2019.

\bibitem{7194625}
S.~Nanz and C.~A. Furia.
\newblock A comparative study of programming languages in rosetta code.
\newblock In {\em 2015 IEEE/ACM 37th IEEE International Conference on Software
  Engineering}, volume~1, pages 778--788, May 2015.

\bibitem{norick2010effects}
B.~Norick, J.~Krohn, E.~Howard, B.~Welna, and C.~Izurieta.
\newblock Effects of the number of developers on code quality in open source
  software: a case study.
\newblock In {\em Proceedings of the 2010 ACM-IEEE International Symposium on
  Empirical Software Engineering and Measurement}, pages 1--1, 2010.

\bibitem{IronTriangle}
R.~Oisen.
\newblock Can project management be defined?
\newblock {\em Project Management Quarterly}, 2(1):12--14, 1971.

\bibitem{Oliveira2020TLProd}
E.~Oliveira, E.~Fernandes, I.~Steinmacher, M.~Cristo, T.~Conte, and A.~Garcia.
\newblock Code and commit metrics of developer productivity: a study on team
  leaders perceptions.
\newblock {\em Empirical Software Engineering}, 04 2020.

\bibitem{876288}
L.~Prechelt.
\newblock An empirical comparison of seven programming languages.
\newblock {\em Computer}, 33(10):23--29, Oct 2000.

\bibitem{6606589}
F.~Rahman and P.~Devanbu.
\newblock How, and why, process metrics are better.
\newblock In {\em 2013 35th International Conference on Software Engineering
  (ICSE)}, pages 432--441, May 2013.

\bibitem{Rahman2011BugCacheFI}
F.~Rahman, D.~Posnett, A.~Hindle, E.~T. Barr, and P.~T. Devanbu.
\newblock Bugcache for inspections: hit or miss?
\newblock In {\em SIGSOFT FSE}, 2011.

\bibitem{NIPS2016_6523}
A.~J. Ratner, C.~M. De~Sa, S.~Wu, D.~Selsam, and C.~R\'{e}.
\newblock Data programming: Creating large training sets, quickly.
\newblock In D.~D. Lee, M.~Sugiyama, U.~V. Luxburg, I.~Guyon, and R.~Garnett,
  editors, {\em Advances in Neural Information Processing Systems 29}, pages
  3567--3575. Curran Associates, Inc., 2016.

\bibitem{Ray:2014:LSS:2635868.2635922}
B.~Ray, D.~Posnett, V.~Filkov, and P.~Devanbu.
\newblock A large scale study of programming languages and code quality in
  github.
\newblock In {\em Proceedings of the 22Nd ACM SIGSOFT International Symposium
  on Foundations of Software Engineering}, FSE 2014, pages 155--165, New York,
  NY, USA, 2014. ACM.

\bibitem{LinusRule}
E.~Raymond.
\newblock The cathedral and the bazaar.
\newblock {\em First Monday}, 3(3), 1998.

\bibitem{Reddivari2019SoftwareQP}
S.~Reddivari and J.~Raman.
\newblock Software quality prediction: An investigation based on machine
  learning.
\newblock {\em 2019 IEEE 20th International Conference on Information Reuse and
  Integration for Data Science (IRI)}, pages 115--122, 2019.

\bibitem{10.2307/1990888}
H.~G. Rice.
\newblock Classes of recursively enumerable sets and their decision problems.
\newblock {\em Transactions of the American Mathematical Society},
  74(2):358--366, 1953.

\bibitem{rosenberg1997some}
J.~Rosenberg.
\newblock Some misconceptions about lines of code.
\newblock In {\em Proceedings fourth international software metrics symposium},
  pages 137--142. IEEE, 1997.

\bibitem{Sackman:1968:EES:362851.362858}
H.~Sackman, W.~J. Erikson, and E.~E. Grant.
\newblock Exploratory experimental studies comparing online and offline
  programming performance.
\newblock {\em Commun. ACM}, 11(1):3--11, Jan 1968.

\bibitem{schach03b}
S.~R. Schach, B.~Jin, L.~Yu, G.~Z. Heller, and J.~Offutt.
\newblock Determining the distribution of maintenance categories: Survey versus
  measurement.
\newblock {\em Empirical Softw.\ Eng.}, 8(4):351--365, Dec 2003.

\bibitem{schneidewind02}
N.~F. Schneidewind.
\newblock Body of knowledge for software quality measurement.
\newblock {\em Computer}, 35(2):77--83, Feb 2002.

\bibitem{Settles10activelearning}
B.~Settles.
\newblock Active learning literature survey.
\newblock Technical report, University of Wisconsin–Madison, 2010.

\bibitem{shepperd88}
M.~Shepperd.
\newblock A critique of cyclomatic complexity as a software metric.
\newblock {\em Software Engineering J.}, 3(2):30--36, Mar 1988.

\bibitem{Shihab:2012:ISR:2393596.2393670}
E.~Shihab, A.~E. Hassan, B.~Adams, and Z.~M. Jiang.
\newblock An industrial study on the risk of software changes.
\newblock In {\em Proceedings of the ACM SIGSOFT 20th International Symposium
  on the Foundations of Software Engineering}, FSE '12, pages 62:1--62:11, New
  York, NY, USA, 2012. ACM.

\bibitem{shrikanth2019assessing}
N.~C. Shrikanth and T.~Menzies.
\newblock Assessing practitioner beliefs about software defect prediction.
\newblock In {\em Intl.\ Conf.\ Softw.\ Eng.}, number~42, May 2020.

\bibitem{nc2020assessing}
N.~C. Shrikanth, W.~Nichols, F.~M. Fahid, and T.~Menzies.
\newblock Assessing practitioner beliefs about software engineering.
\newblock arXiv:2006.05060, June 2020.

\bibitem{Sliwerski:2005:CIF:1082983.1083147}
J.~\'{S}liwerski, T.~Zimmermann, and A.~Zeller.
\newblock When do changes induce fixes?
\newblock {\em SIGSOFT Softw. Eng. Notes}, 30(4):1--5, May 2005.

\bibitem{Swanson:1976:DM:800253.807723}
E.~B. Swanson.
\newblock The dimensions of maintenance.
\newblock In {\em Proceedings of the 2Nd International Conference on Software
  Engineering}, ICSE '76, pages 492--497, Los Alamitos, CA, USA, 1976. IEEE
  Computer Society Press.

\bibitem{6676898}
S.~E.~S. {Taba}, F.~{Khomh}, Y.~{Zou}, A.~E. {Hassan}, and M.~{Nagappan}.
\newblock Predicting bugs using antipatterns.
\newblock In {\em 2013 IEEE International Conference on Software Maintenance},
  pages 270--279, 2013.

\bibitem{tom2013exploration}
E.~Tom, A.~Aurum, and R.~Vidgen.
\newblock An exploration of technical debt.
\newblock {\em Journal of Systems and Software}, 86(6):1498--1516, 2013.

\bibitem{1173068}
E.~{van Emden} and L.~{Moonen}.
\newblock Java quality assurance by detecting code smells.
\newblock In {\em Ninth Working Conference on Reverse Engineering, 2002.
  Proceedings.}, pages 97--106, Nov 2002.

\bibitem{van2002java}
E.~Van~Emden and L.~Moonen.
\newblock Java quality assurance by detecting code smells.
\newblock In {\em Ninth Working Conference on Reverse Engineering, 2002.
  Proceedings.}, pages 97--106. IEEE, 2002.

\bibitem{Vasilescu:2015:QPO:2786805.2786850}
B.~Vasilescu, Y.~Yu, H.~Wang, P.~Devanbu, and V.~Filkov.
\newblock Quality and productivity outcomes relating to continuous integration
  in github.
\newblock In {\em Proceedings of the 2015 10th Joint Meeting on Foundations of
  Software Engineering}, ESEC/FSE 2015, pages 805--816, New York, NY, USA,
  2015. ACM.

\bibitem{Walkinshaw:2018:FRD:3239235.3239244}
N.~Walkinshaw and L.~Minku.
\newblock Are 20\% of files responsible for 80\% of defects?
\newblock In {\em Proceedings of the 12th ACM/IEEE International Symposium on
  Empirical Software Engineering and Measurement}, ESEM '18, pages 2:1--2:10,
  New York, NY, USA, 2018. ACM.

\bibitem{Weyuker08Spoil}
E.~Weyuker, T.~Ostrand, and R.~Bell.
\newblock Do too many cooks spoil the broth? using the number of developers to
  enhance defect prediction models.
\newblock {\em Empirical Software Engineering}, 13:539--559, 10 2008.

\bibitem{10.5555/548833}
L.~Williams and R.~Kessler.
\newblock {\em Pair Programming Illuminated}.
\newblock Addison-Wesley Longman Publishing Co., Inc., USA, 2002.

\bibitem{544240}
A.~{Wood}.
\newblock Predicting software reliability.
\newblock {\em Computer}, 29(11):69--77, Nov 1996.

\bibitem{Wright00Satisfaction}
T.~A. Wright and R.~Cropanzano.
\newblock Psychological well-being and job satisfaction as predictors of job
  performance.
\newblock {\em Journal of Occupational Health Psychology}, 5:84--94, 2000.

\bibitem{1701965}
S.~{Yamada} and S.~{Osaki}.
\newblock Software reliability growth modeling: Models and applications.
\newblock {\em IEEE Transactions on Software Engineering},
  SE-11(12):1431--1437, Dec 1985.

\bibitem{yamashita2012code}
A.~Yamashita and L.~Moonen.
\newblock Do code smells reflect important maintainability aspects?
\newblock In {\em 2012 28th IEEE international conference on software
  maintenance (ICSM)}, pages 306--315. IEEE, 2012.

\bibitem{1231213}
T.~Zimmermann, S.~Diehl, and A.~Zeller.
\newblock How history justifies system architecture (or not).
\newblock In {\em Sixth International Workshop on Principles of Software
  Evolution, 2003. Proceedings.}, pages 73--83, Sept 2003.

\end{thebibliography}
